\documentclass[letterpaper,twocolumn,10pt]{sig-alternate-10pt}
\usepackage{times}
\usepackage{amsfonts}
\usepackage[T1]{fontenc}
\usepackage{textcomp}

\usepackage[usenames,dvipsnames]{xcolor}
\usepackage[pdfstartview=FitH, bookmarksnumbered=true, bookmarksopen=true, colorlinks=true, citecolor=blue, colorlinks=blue, linkcolor= blue, urlcolor=blue]{hyperref}
\usepackage{soul}
\usepackage{url}
\usepackage[hang,scriptsize,tight,nooneline]{subfigure}

\usepackage{xspace}
\usepackage{flushend}
\usepackage{enumitem}
\newcommand{\ie}{{\it i.e.},\xspace}
\newcommand{\cut}[1]{}
\newcommand{\cutatlastminute}[1]{}

\newcommand{\paragraphb}[1]{\vspace{0.04in}\noindent{\bf #1} }

\newcommand{\ankitnew}{\textcolor{black}}

\newcommand{\nnn}{\textcolor{black}}

\setstcolor{red}
\begin{document}
%\conferenceinfo{HotNets '14,} { October 27--28, 2014, Los Angeles, CA,
%USA.}
%\CopyrightYear{2014}
%\crdata{Copyright held by the owner/author(s). Publication rights licensed to ACM. 
%ACM 978-1-4503-3256-9/14/10 ... \$15.00 \\
%http://dx.doi.org/10.1145/2670518.2673876}
%\clubpenalty=10000
%\widowpenalty = 10000
%%%%%%%%%%%%%%%%%%%%%%%%%%%%%%%%%%%%%%%%%%%%%%%%%%%%%%%%%%%%%%%%%%%%%%%%%%%%%%%%
%\pagenumbering{gobble}

\title{Towards a Speed of Light Internet}

\author{
Ankit Singla$^\dag$, Balakrishnan Chandrasekaran$^\sharp$, P. Brighten Godfrey$^\dag$, Bruce Maggs$^\sharp$$^*$\\
$^\dag$University of Illinois at Urbana--Champaign, $^\sharp$Duke University, $^*$Akamai\\
$^\dag$\{singla2, pbg\}@illinois.edu, $^\sharp$\{balac, bmm\}@cs.duke.edu
} 
\maketitle
\abstract{
%	For many Internet services, reducing latency improves the user experience and increases revenue for the service provider. While
	
In principle, a network can transfer data at nearly the speed of
light.  Today's Internet, however, is much slower: our measurements
show that latencies are typically more than one, and often more than
two orders of magnitude larger than the lower bound implied by the
speed of light.  Closing this gap would not only add value to today's
Internet applications, but might also open the door to exciting new
applications. Thus, we propose a grand challenge for the networking
research community: building a speed-of-light Internet. Towards
addressing this goal, we begin by investigating the causes of latency
inflation in the Internet across the network stack. Our analysis
reveals that while protocol overheads, which have dominated the
community's attention, \ankitnew{are indeed important, infrastructural inefficiencies are a significant and under-explored problem.} Thus,
we propose a radical, yet surprisingly low-cost approach to mitigating
latency inflation at the lowest layers and building a nearly
speed-of-light Internet infrastructure.}

%\category{C.2.1}{Computer-Communication Networks}{Network Architecture and Design}
%\category{C.2.5}{Computer-Communication Networks}{Local and Wide-Area Networks}[Internet]
%\category{C.4}{Performance of Systems}{Performance attributes}

%\terms{Measurement; Design; Performance}

\section{Introduction}

Reducing latency across the Internet is of immense value ---
measurements and analysis by Internet giants have shown that shaving a
few hundred milliseconds from the time for a transaction can translate
into millions of dollars. For Amazon, a $100$ms latency penalty
implies a $1\%$ sales loss~\cite{amazonLatency}; for Google, an
additional delay of $400$ms in search responses reduces search volume
by $0.74\%$; and for Bing, $500$ms of latency decreases revenue per
user by $1.2\%$~\cite{brutlag,schurman}. Undercutting a competitor's
latency by as little as $250$ms is considered a competitive
advantage~\cite{msMatter} in the industry.  Even more crucially, these
numbers underscore that latency is a key determinant of user
experience.

While latency reductions of a few hundred milliseconds are valuable,
we take the position that the networking community should pursue a
much more ambitious goal: cutting Internet latencies to close to the
limiting physical constraint, the speed of light, roughly one to two
orders of magnitude faster than today.  What would such a drastic
reduction in Internet latency mean, and why is it worth pursuing?
Beyond the obvious gains in performance and value for today's
applications, such a technological leap could have truly
transformative impact.  A speed-of-light Internet may help realize the
full potential of certain applications that have so far been limited
to the laboratory \ankitnew{such as tele-immersion}. For some
applications, such as massive multi-player online games, the size of
the user community reachable within a latency bound plays an important
role in user interest and adoption and, as we shall see later, linear
decreases in communication latency result in super-linear growth in
community size. Low latencies on the order of a few tens of
milliseconds also open up the possibility of \emph{instant response},
where users are unable to perceive any lag between requesting a page
and seeing it rendered in their browsers. Such an elimination of wait
time would be an important threshold in user experience. A
speed-of-light Internet can also be expected to spur the development
of new and creative applications. The creators of the Internet, after
all, did not envision the myriad ways in which it is used today.

But the Internet's speed is quite far from the speed of light.  As we
show later, the fetch time from a set of generally well-connected
clients for just the HTML document of the index pages of
popular Web sites is, in the median, $35$ times the round-trip
speed-of-light latency. In the $80^{th}$ percentile it is more than
$100\times$ slower. Given the promise a speed-of-light Internet holds,
\emph{why are we so far from the speed of light?}

While ISPs compete primarily on the basis of peak bandwidth offered,
bandwidth is not the issue.  Bandwidth improvements are also
necessary, but bandwidth is no longer the bottleneck for a significant
fraction of the population: for instance, the average Internet
connection speed in the US is $11.5$Mbps~\cite{akamaiState}, while the
effect of increasing bandwidth on page load time is small beyond as
little as $5$Mbps~\cite{ilya-latency}.  Projects like Google
Fiber~\cite{googleFiber} and other fiber-to-the-home efforts by ISPs
are further improving bandwidth. On the other hand, it has been noted
in a variety of contexts from CPUs, to disks, to networks, that
`latency lags bandwidth', and reducing latency is a more difficult
problem~\cite{patterson2004latency}.

% In this work, we motivate the ambitious agenda of building a
% speed-of-light Internet, quantify the causes of Internet latency
% inflation, and put forth a radical proposal aimed at building
% speed-of-light Internet infrastructure. Our contributions are as
% follows:

If bandwidth isn't the culprit, then what is?! To identify the causes
of latency inflation, we use two large datasets: $2.9$ million
measurements $28$,$000$ Web page downloads from $186$ clients to
servers in $103$ countries; and $2.4$ million latency measurements
between \ankitnew{servers at a large CDN provider} and end-users
($1.7$ million host pairs). We augment this data with IP geolocation
data from five geolocation services. Our analysis of this data breaks
down Internet latency inflation across the network stack, from the
physical network infrastructure through the transport layer. In line
with the community's understanding, DNS, TCP's three-way handshake,
and TCP slow-start are \ankitnew{all important factors in latency inflation.
Also significant however, are the Internet's infrastructural
deficiencies.} Apart from quantifying the contributions of various
protocol factors to latency inflation, a key contribution of this work
is putting latency inflation at the lowest layers in context --- as we
discuss in \S\ref{sec:whyslow}, \ankitnew{the effect of improvements
  at the lowest layers is multiplicative.} We consider this an
under-appreciated piece of the latency puzzle.

Further, while networking research has largely taken
infrastructure as a given and focused on latency improvements in
network protocols, we show that infrastructural impediments to latency
can be addressed surprisingly cheaply and with immediate
payoff. \cut{Given the large contribution of infrastructural latency
  inflation to poor latencies on the Internet, this is a massive
  opportunity.}We propose a radical approach to building a parallel,
nearly speed-of-light Internet infrastructure to augment our current
high-capacity Internet backbone (\S\ref{sec:proposal}). We analyze the
cost, coverage, and effectiveness of our approach through a case study
in designing such an infrastructure for the contiguous United States. Our
key finding is that a nearly speed-of-light Internet infrastructure
connecting $85\%$ of the US population can be built at the cost of a
few hundred million dollars. In the context of Internet
infrastructure, where the cost of \emph{individual} submarine cables
can be many times larger~\cite{arcticCable}, this is surprisingly
cheap. Ironically, in this case, infrastructure might be an easier
avenue for rapid improvement than network protocols!

However, as our measurements reveal, achieving speed-of-light
latencies also depends on faster protocols. While the
thrust of this paper is on tackling latency inflation at the lowest
layers, recent advances in protocol research, which we discuss in
\S\ref{subsec:protocols}, provide us with many possible solutions that
may be put together to build a speed-of-light Internet.

In summary, we propose a `speed-of-light Internet' as a grand
challenge for the networking community, quantify the factors that
contribute to large latencies today, and propose a radical, but
viable, approach to closing the latency gap.

\section{The need for speed}
\label{sec:nfs}

\begin{figure*}
\centering
\hspace{-34pt}
\subfigure[]{ \label{fig:commSize}\includegraphics[width=2.30in]{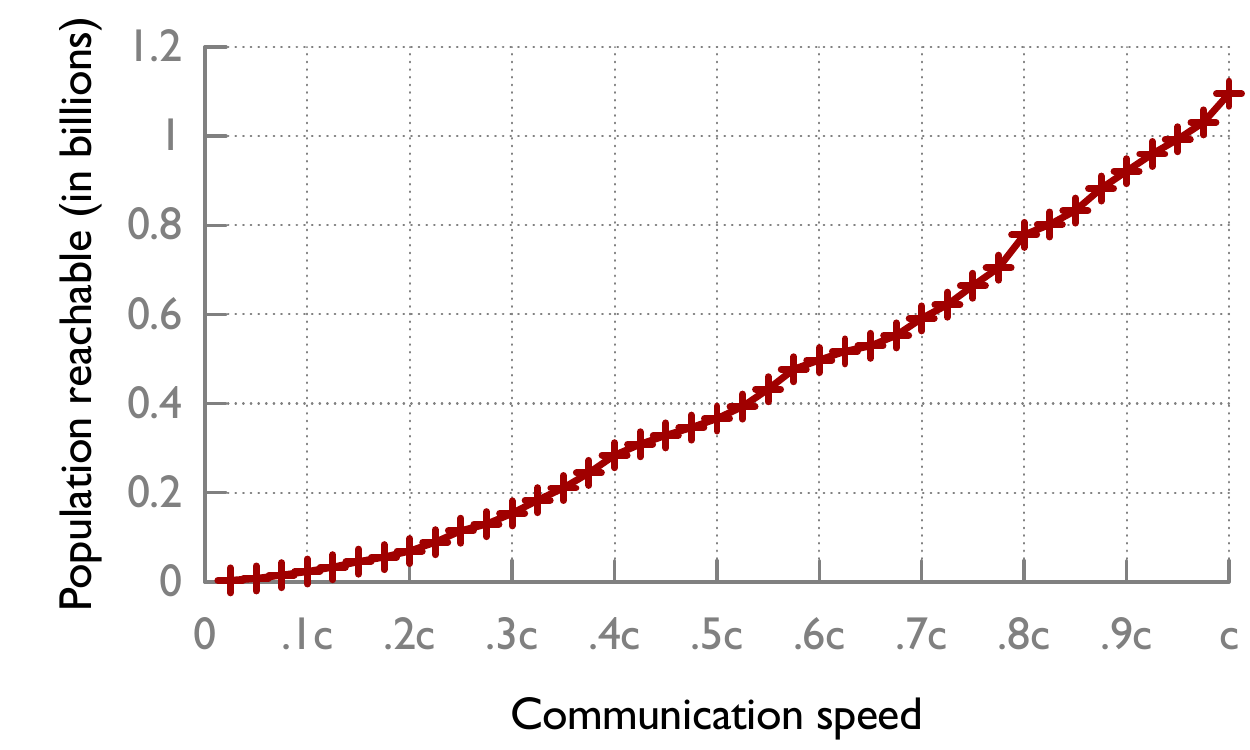}}
\subfigure[]{ \label{fig:numDcns}\includegraphics[width=2.30in]{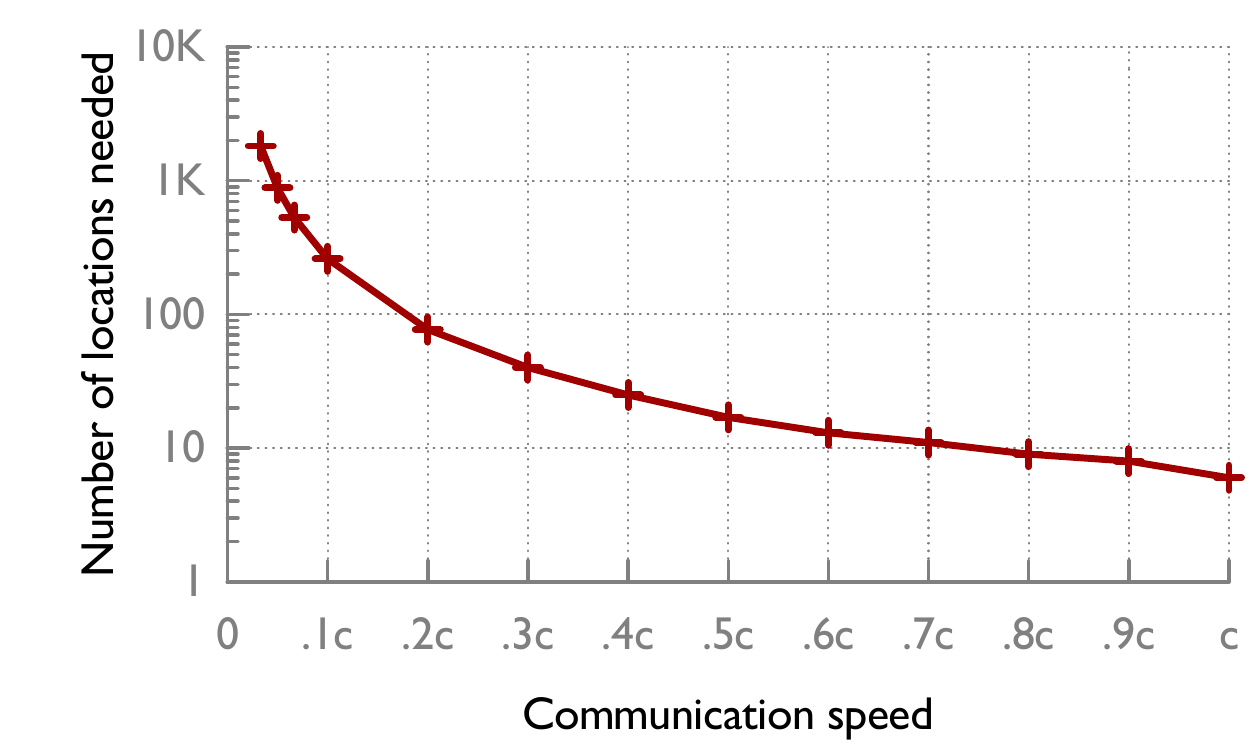}}
\subfigure[]{ \label{fig:tradeoff}\includegraphics[width=2.30in]{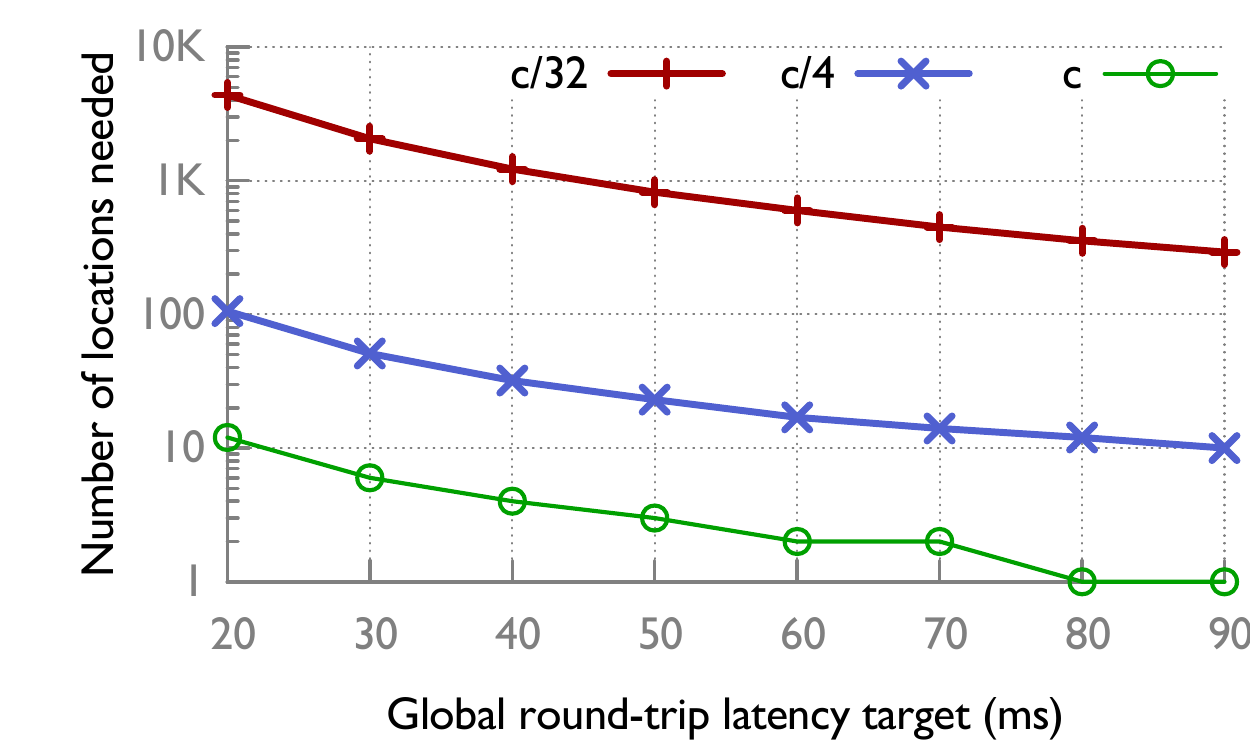}}
\hspace{-20pt}
\vspace{-8pt}
\caption{\small \em The impact of communication speed on computing and people. With increasing communication speed: (a) the population within $30$ms round-trip time grows super-linearly;
%For each of $200$ world capital cities, we calculate the number of people within $30$ms rount-trip time at a certain communication speed ($x$-axis); the $y$-axis displays the median of these numbers. 
(b) the number of locations (e.g. data centers or CDN nodes) needed for global $30$ms reachability from at least one location falls super-linearly; and (c) the tradeoff between the global latency target
 and the number of locations required to meet it improves.}
 \vspace{-6pt}
 \end{figure*}
 
%Where even a few hundred milliseconds translate to millions of dollars for Web services and their users, 
A speed-of-light Internet would be an advance with tremendous impact. It would enhance user satisfaction with Web applications, as well as voice and video communication. The gaming industry, where latencies larger than $50$ms can hurt gameplay~\cite{gamingLatency}, would also benefit. But beyond the promise of these valuable improvements, a speed-of-light Internet could fundamentally transform the computing landscape. 

\paragraphb{New applications.}
One of computing's natural, yet unrealized goals is to create a convincing experience of joining two distant locations. \ankitnew{In fact, \emph{tele-immersion} topped the list of `Killer Apps in the Gigabit Age' themes in a Pew survey of $1400$$+$ experts\footnote{Including Vint Cerf, Bob Briscoe, and David Clark.}~\cite{pewsurveyGigabit}. 
%For locations that are too distant (such as, say, Beijing and Buenos Aires) the speed of light barrier may prohibit such a bridged experience. Nevertheless, a speed-of-light Internet would enable large swathes of the globe to communicate as such. 
%A speed-of-light Internet can finally enable us to achieve this goal.
Applications like tele-immersion and remote collaborative music performance are hampered today by poor Internet latencies. For instance, latencies above $50$ms, make remote collaboration on music difficult~\cite{chewMusic}. Convincing virtual reality immersion necessitates a latency of less than $20$ms~\cite{vrImmersion}, and a similar limit likely applies to immersion in remote, real-world environments. Note that VR immersion is vastly different from what is marketed as telepresence today, where communicating parties interact in a very static environment.} A speed-of-light Internet could move such applications from their limited experimental scope, to ubiquity. And perhaps we will be surprised by the creative new applications that evolve in that environment\footnote{``New capabilities emerge just by virtue of having smart people with access to state-of-the-art technology.'' --- Bob Kahn}.

\paragraphb{Illusion of instant response.}
A speed-of-light Internet can realize the possibility of \emph{instant response}. The human visual system cannot correctly order visual events separated by less than about $30$ms~\cite{visualThresh}. Thus, if responses over the Internet were received within $30$ms of the requests, we would achieve the illusion of instant response\footnote{This is a convenient benchmark number, but the exact number will vary depending on the scenario. For a $30$ms response time, the Internet will actually need to be a little faster because of server-side request processing time, screen refresh delay, etc. And the `instant response' threshold will differ somewhat for audio vs. visual applications.}. A (perceived) zero wait-time for Internet services would greatly improve user experience and allow for richer interaction. Immense resources, both computational and human, would become ``instantly'' available over a speed-of-light Internet.

\paragraphb{The Internet of Things.} Tens of billions of devices (excluding traditional personal computing devices) are expected to be on the Internet of Things by $2020$~\cite{gartnerIOT,abiIOT}. For some of these devices, latency may not matter much -- for a `smart' thermostat to respond to temperature settings within several seconds is fine. However, other `smart things' may be intended to facilitate active interaction with people, for example, shoppers using their wearable electronics to interact with merchandise that has smart tags. Such an augmented reality may depend on low-latency access to information over the Internet. \ankitnew{Latencies close to the limits of human perception would make these interactions seamless and natural.}

\paragraphb{Super-linear community size.} Many applications require that the connected users be reachable within a certain latency threshold, such as $30$ms round-trip for instant response, or perhaps $50$ms for a massive multi-player game. The value of low latency is magnified by the fact that \emph{the size of the available user community is a superlinear function of network speed}. The area on the Earth's surface reachable within a given latency grows nearly\footnote{Because the Earth is a sphere, not a plane.} quadratically in latency. Using population density data\footnote{Throughout, we use population estimates for $2010$~\cite{populationData}.} reveals somewhat slower, but still super-linear growth. We measured the number of people within a $30$ms RTT from $200$ World capital cities at various communication speeds. Fig.~\ref{fig:commSize} shows the median (across cities) of the population reached. If Internet latencies were $20\times$ worse than $c$-latency ($x$-axis=$0.05c$), we could reach $7.5$ million people ``instantly''. A \textbf{10}$\times$ latency improvement ($x$-axis=$0.5c$) would increase that community size by \textbf{49}$\times$. Therefore, the value of latency improvement is magnified, perhaps pushing some applications to reach critical mass.

\paragraphb{Cloud computing and thin clients.}
Another potential effect of a speedier Internet is further centralization of compute resources. Google and VMware are already jointly working towards the thin client model through virtualization~\cite{googVMware}. Currently, their Desktop-as-a-Service offering is targeted at businesses, with the customer centralizing most compute and data in a cluster, and deploying cheaper hardware as workstations. A major difficulty with extending this model to personal computing today is the much larger latency involved in reaching home users. Likewise, in the mobile space, there is interest in offloading some compute to the cloud, thereby exploiting data and computational resources unavailable on user devices~\cite{cuervo}. 
% --- applications such as Apple's Siri~\cite{siriOffload} and Google Now~\cite{googNow} offload some processing from their respective mobile platforms to the cloud. These applications make use of data and computational resources unavailable on user devices, but still must provide low response times that enable interactivity. 
As prior work~\cite{ha2013impact} has argued, however, to achieve highly responsive performance from such applications would today require the presence of a large number of data center facilities. With a speedier Internet, the `thin client' model becomes plausible for both desktop and mobile computing with far fewer installations. For instance, if the Internet operated at half the speed of light, almost all of the contiguous US could be served instantly from just one location. Fig.~\ref{fig:numDcns} shows the number of locations needed for $99\%$ of the world's population to be able to instantly reach at least one location --- as we decrease Internet latency, the number of facilities required falls drastically, down to only $6$ locations with global speed-of-light connectivity. (These numbers were estimated using a heuristic placement algorithm and could possibly be improved upon.) This result is closely related to that in Fig.~\ref{fig:commSize} --- with increasing communication speed (which, given a latency bound, determines a reachable radius), the population reachable from a center grows super-linearly, and the number of centers needed to cover the entire population falls super-linearly.

\paragraphb{Better geolocation.} As latency gets closer to the speed of light, latency-based geolocation gets better, and in the extreme case of exact $c$-latency, location can be precisely triangulated.\cut{A speed-of-light Internet will also enable accurate geolocation. Consider, for example, a client located a certain distance away from a server. Over a speed-of-light Internet, the round-trip latency will reveal to the server precisely this distance. (As Internet latency inflation and variations in it increase, the server's estimates lose precision.) Simple techniques such as triangulation would yield highly accurate geolocation over a speed-of-light Internet.} While better geolocation provides benefits such as better targeting of services and matching with nearby servers, it also has other implications, such as for privacy.

\paragraphb{Don't CDNs solve the latency problem?} Content distribution networks cut latency by placing a large number of replicas of content across the globe, so that for most customers, some replica is nearby. However, this approach has its limitations. First, some resources simply cannot be replicated or moved, such as people --- so CDNs are not relevant for all communication. Second, CDNs today are an expensive option, available only to larger Internet companies. \ankitnew{A speedier Internet would significantly cut costs for CDNs, putting them within reach of a larger spectrum of Web service providers.} CDNs make a tradeoff between costs (determined, in part, by the number of infrastructure locations), and latency targets. For any latency target a CDN desires to achieve globally, given the Internet's communication latency, a certain minimum number of locations are required. Speeding up the Internet improves this entire tradeoff curve. This improvement is shown in Fig.~\ref{fig:tradeoff}, where we estimate (using our random placement heuristic) the number of locations required to achieve different latency targets for different Internet communication speeds\footnote{Per our measurements in \S\ref{sec:tooslow}, $\frac{c}{32}$ is close to the median speed of fetching just the HTML for the landing pages of popular websites today, and $\frac{c}{4}$ is close to the median ping speed.}: $\frac{c}{32}$, $\frac{c}{4}$, and $c$. As is clear from these results, while CDNs will still be necessary to hit global latency targets of a few tens of milliseconds, the amount of infrastructure they require to do so will fall drastically with a speedier Internet.

\section{The Internet is too slow}
%\label{sec:experiment}
\label{sec:tooslow}
%\vspace{-7pt}
%\subsection{How slow is the Internet?}
%\label{subsec:howslow}

\begin{figure}
\centering
\includegraphics[width=2.5in]{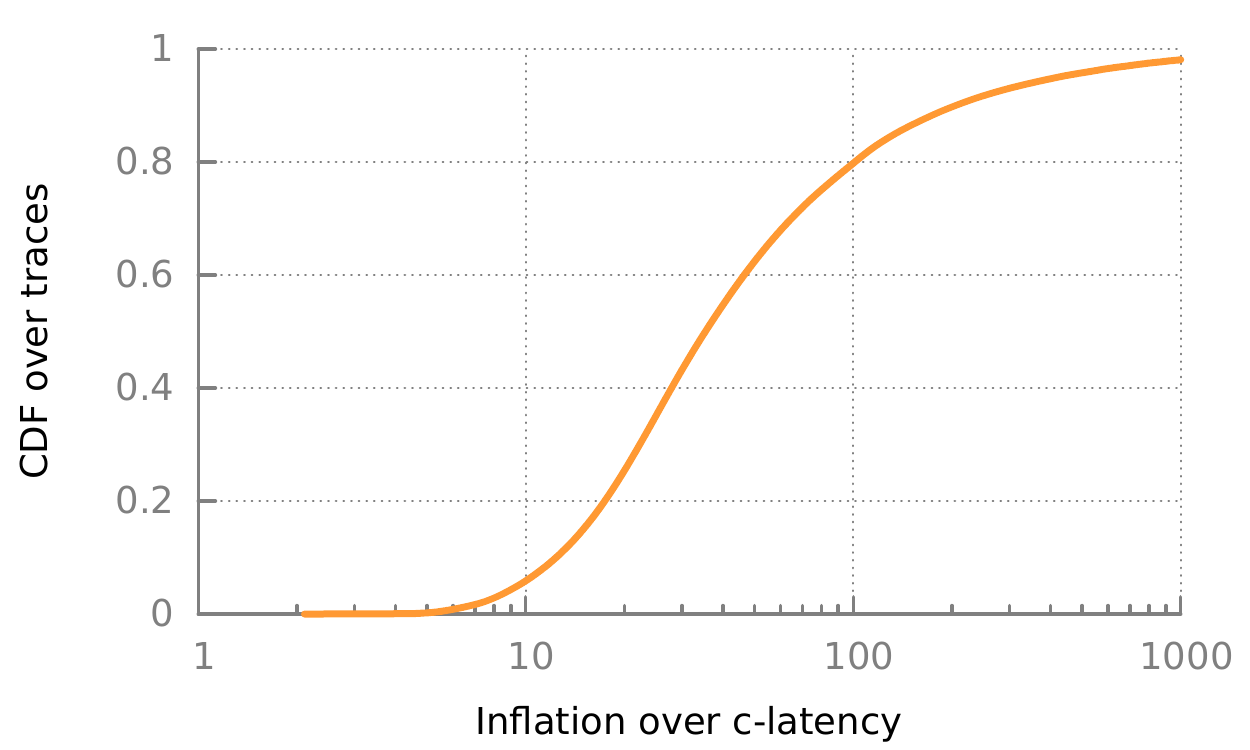}
\vspace{-8pt}
\caption{\small \em Fetch time of just the HTML of the landing pages of popular Web sites in terms of inflation over the speed of light.}
%In the median, fetch time is $35.4\times$ slower.}
\label{fig:howslow}
\end{figure}

We fetched just the HTML for the landing pages of $28$,$000$ popular Web sites
from $186$ PlanetLab locations using cURL~\cite{curl}. We pooled Alexa's~\cite{alexaByCountry}
top $500$ Web sites from each of $103$ countries and used the unique URLs. We followed
redirects on each URL, and recorded the final URL for use in our measurements. In our experiments,
we ignored any URLs that still caused redirects. The results presented here exclude data for the few hundreds of Web sites in our sample that use TLS/SSL; an analysis accounting for the latency cost of establishing secure connections is left to future work.

For each connection, we geolocated the Web server using commercial geolocation services, and computed the
time it would take for light to travel round-trip along the shortest path between the same
end-points, \ie the $c$-latency\footnote{We have ground-truth geolocation for PlanetLab nodes --- while the PlanetLab API yields incorrect locations for some nodes, these are easy to identify and remove based on simple latency tests.}.
Henceforth, we refer to the ratio of the fetch time to
$c$-latency as the Internet's latency inflation. Fig.~\ref{fig:howslow} shows the CDF of
inflation over $2.9$ million connections. The HTML fetch time is, in the median, $35.4\times$ the
 $c$-latency, while the $80^{th}$ percentile exceeds $100\times$. Thus, the Internet is typically
 more than an order of magnitude slower than the speed of light, and often two orders of magnitude slower. We note that PlanetLab
 nodes are generally well-connected, and latency can be expected to be poorer from the network's
 \emph{true} edge.

\section{Why is the Internet so slow?}
\label{sec:whyslow}

To identify the causes of Internet latency inflation, we break down the fetch time across layers, from inflation in the physical path followed by packets to the TCP transfer time. We first describe the methodology (\S\ref{subsec:methods}) and an overview of the results (\S\ref{subsec:overall}). Next, we discuss the robustness of our results to IP geolocation errors (\S\ref{subsec:geolocation}), and consistency across page fetch sizes (\S\ref{subsec:sizes}) and geographies (\S\ref{subsec:geographies}). In \S\ref{subsec:congestion}, we investigate the role of congestion, and lastly, in \S\ref{subsec:infra}, the impact of infrastructural deficiencies.

\subsection{Methodology}
\label{subsec:methods}
We use cURL to obtain the time for DNS resolution, TCP handshake, TCP data transfer, and total fetch time for each connection. The TCP handshake is measured as the time between cURL sending the \texttt{SYN} and receiving the \texttt{SYN-ACK}. The TCP transfer time is measured as the time from cURL's receipt of the first byte to the receipt of the last byte. We separately account for the time between cURL sending the data request
%(immediately following the handshake \texttt{ACK})
and the receipt of the first byte as `request-response' time; this typically comprises one RTT and any processing time at the Web server. For each connection, we also run a traceroute from the client PlanetLab node to the Web server. We then geolocate each router in the traceroute, and connect successive routers with the shortest paths on the Earth's
surface as an approximation for the route the packets follow. We compute the roundtrip latency at the speed of light in fiber along this approximate path, and refer to it as the `router-path latency'. From each client, we also run $30$ successive pings to each Web server, and record the minimum and median across these ping times. We normalize each of these latency components by the $c$-latency between the respective connection's end-points.

\begin{figure}
\centering
\includegraphics[width=3in, keepaspectratio]{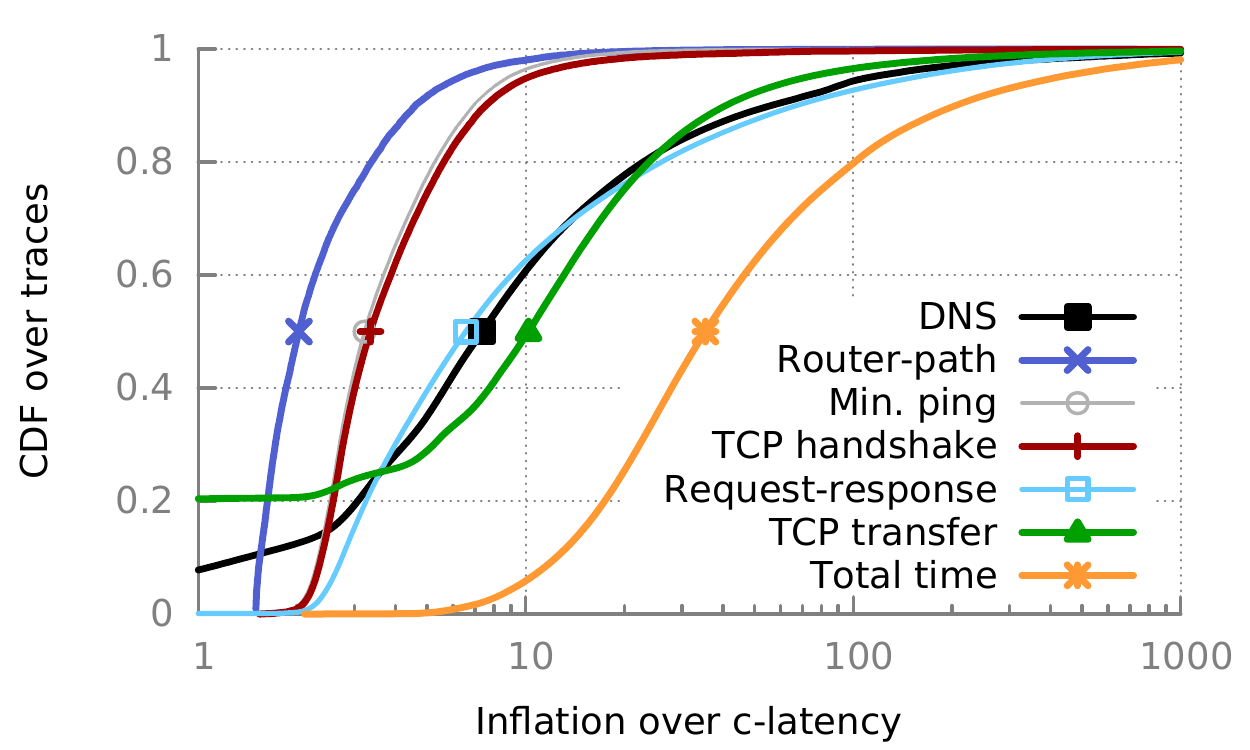}
\vspace{-8pt}
\caption{\small \em Various components of latency inflation. The median is marked on each curve for sake of clarity of the legend.}
\label{fig:prelim}
\end{figure}

\begin{figure*}
\centering
\subfigure[Router-path latency inflation]{\label{fig:geo-rp}  \includegraphics[width=2.25in]{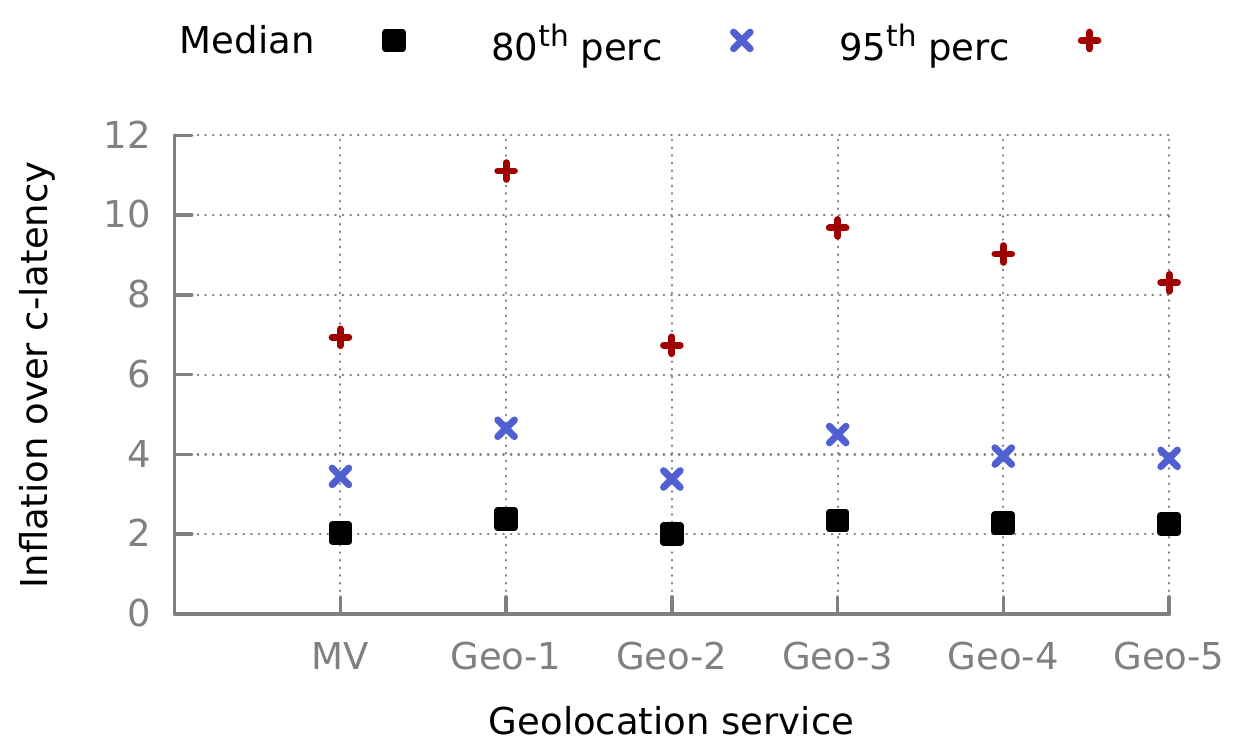}}
\subfigure[Minimum ping latency inflation]{ \label{fig:geo-minp} \includegraphics[width=2.25in]{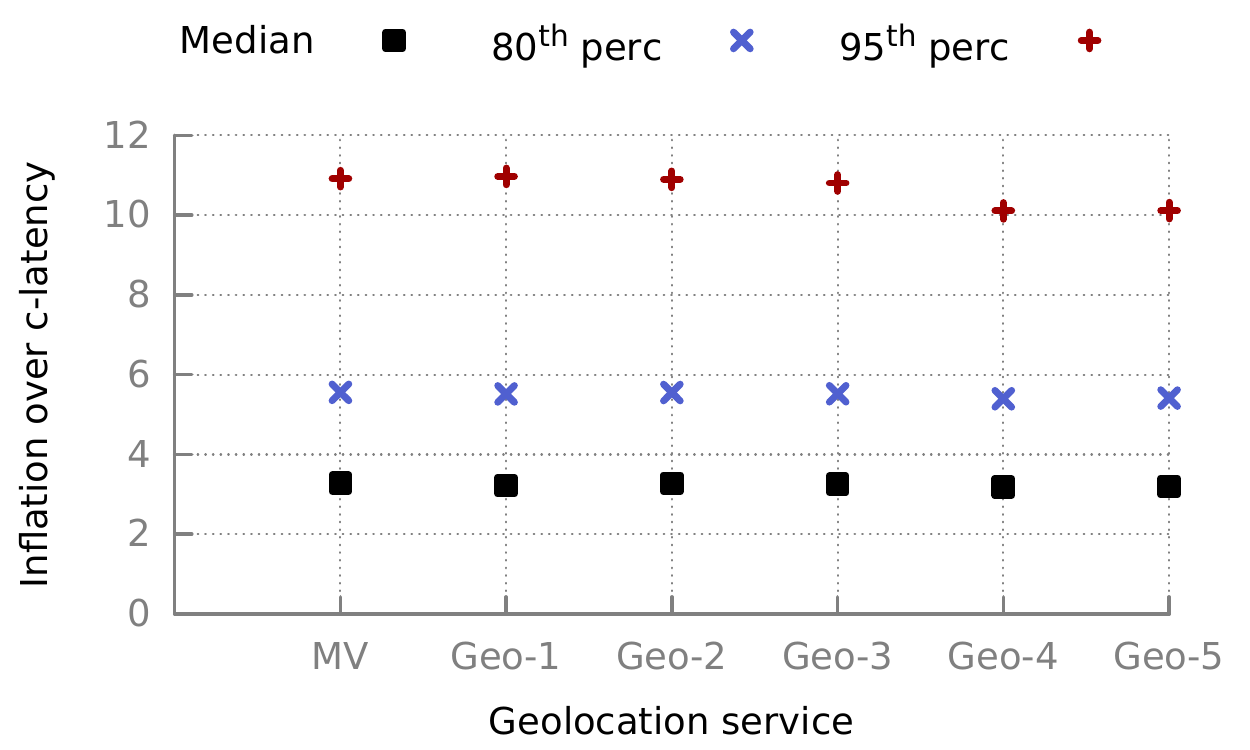}}
\subfigure[Total fetch time inflation]{ \label{fig:geo-total}\includegraphics[width=2.25in]{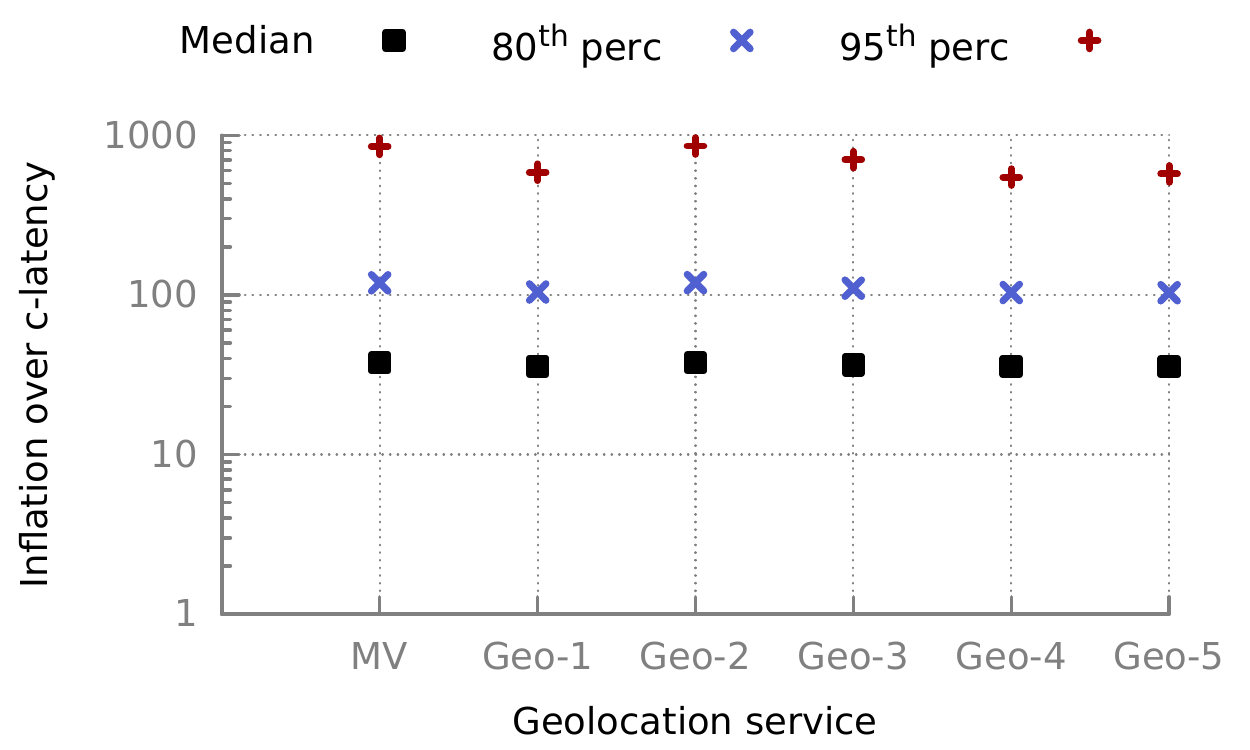}}
\vspace{-8pt}
\caption{\small \em Results for median, $80^{th}$-percentile, and $95^{th}$-percentile of inflation in several metrics, using $5$ different commercial geolocation databases as well as their majority vote (MV).}
%Inflation in minimum ping time and total transfer time are fairly consistent across these results, particularly in the median and the $80^{th}$-percentile, but variation in the router-path latency inflation is larger.}
\label{fig:geolocation}
\vspace{-8pt}
\end{figure*}

%We limit this analysis to roughly one million connections, for which we used cURL to fetch the first $32$KB ($22$ full-sized packets) of data from the Web server\footnote{cURL allows explicit specification of the number of bytes to fetch, but some servers do not honor such a request. Measurements from connections that did not fetch roughly $32$KB were discarded.}.
%We restrict these results to the roughly one million connections that fetch $\sim$$32$KB, are shown in Fig.~\ref{fig:prelim}.

\subsection{Overview of results}
\label{subsec:overall}

Fig.~\ref{fig:prelim} shows the results for all $2.9$ million connections\footnote{Any connections where our data showed obvious anomalies, such as $c$-latency being larger than the minimum ping time due to geolocation errors, were weeded out; $2.9$ million connections survive such checks.}. It is unsurprising that DNS resolutions are faster than $c$-latency $8\%$ of the time --- in these cases, the server happens to be farther than the DNS resolver. (The DNS curve is clipped at the left to more clearly display the other results.) In the median, DNS resolutions are $7.4\times$ inflated over $c$-latency. \cut{In fact, we found that when we consider the connections in the top and bottom $10$ percentiles of total fetch time inflation, DNS plays a significant role -- among the fastest $10\%$ of pages, even the $90^{th}$-percentile DNS inflation is only $5.2\times$, while for the slowest $10\%$ of pages, even the median DNS time is $64\times$ inflated.}

The TCP transfer time shows significant inflation --- $10.2\times$ in the median. \ankitnew{With most pages being tens of KBs, bandwidth is not the problem, but TCP's slow start causes even small data transfers to require several RTTs.} $20\%$ of all pages have transfer times less than the $c$-latency --- this is due to all the data being received in the first TCP window. (Recall that transfer time is the time between cURL receiving the first and the last bytes.) The TCP handshake (counting only the \texttt{SYN} and \texttt{SYN-ACK}) and the minimum ping time are $3.4\times$ and $3.2\times$ inflated in the median.
%\cut{The ping and handshake latencies between the same hosts thus show little variation.}

The request-response time is $6.6\times$ inflated in the median, \ie roughly twice the median round-trip time. \cut{Thus, across our data, the server processing time comprises $\sim$$1$RTT out of the total fetch time in the median. }However, $25\%$ of the connections use less than $10$ms of server processing time (estimated by subtracting one RTT from the request-response time).
%The median $c$-latency, in comparison, is $43$ms.

\begin{figure*}
\centering
\subfigure[]{ \label{fig:fetchsize:dist}    \includegraphics[width=2.25in]{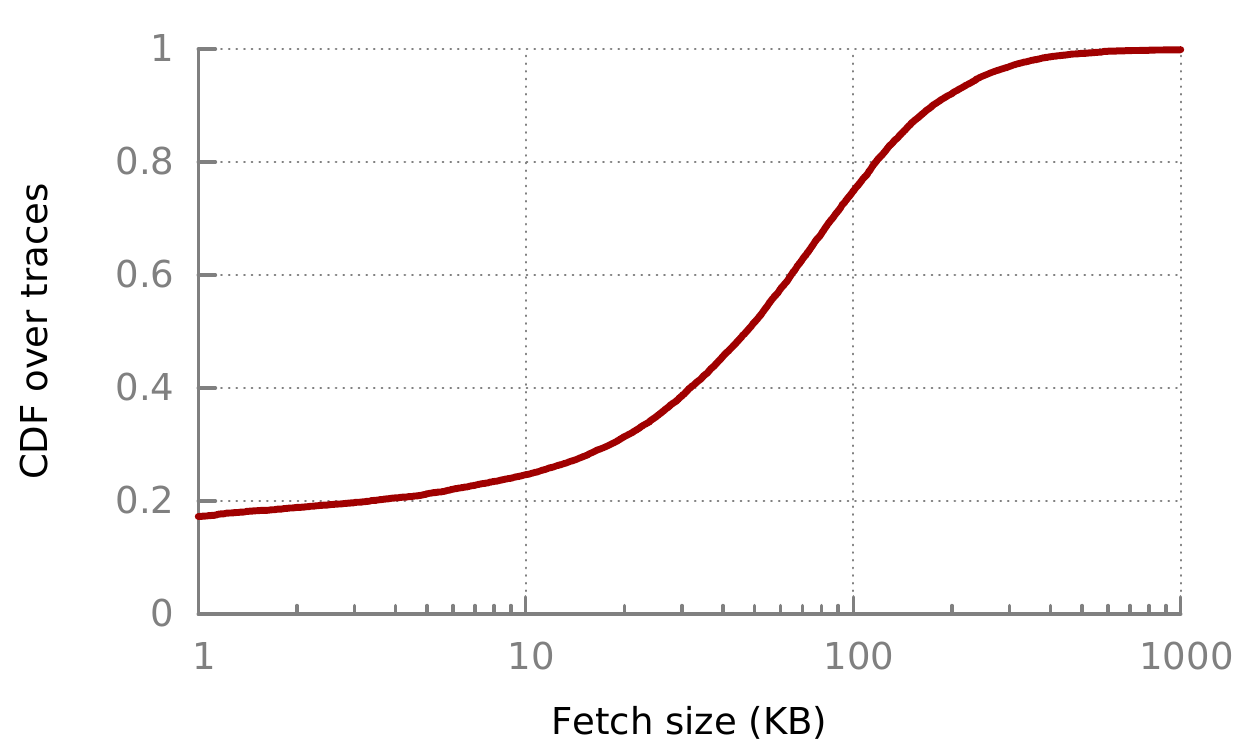}}
%\vspace{-6pt}
\subfigure[]{ \label{fig:fetchsize:sizeInfl}\includegraphics[width=2.25in]{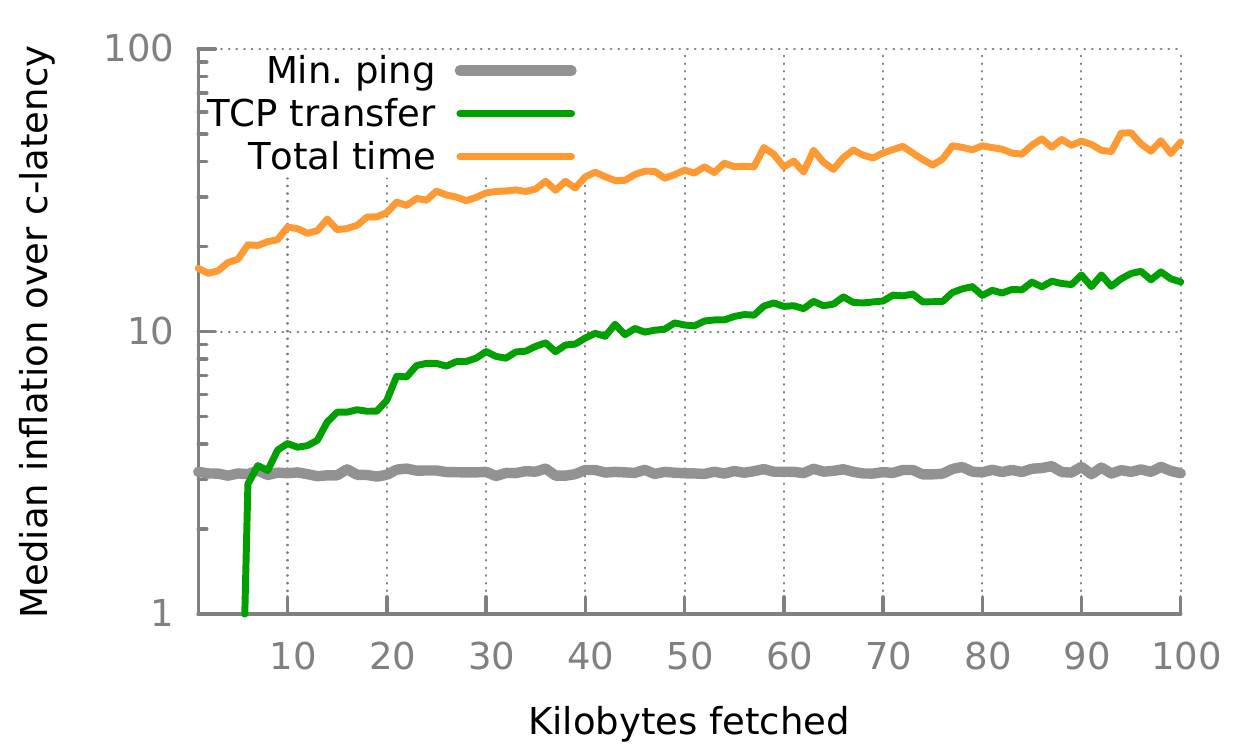}}
%\vspace{-6pt}
\subfigure[]{ \label{fig:fetchsize:medsize} \includegraphics[width=2.25in]{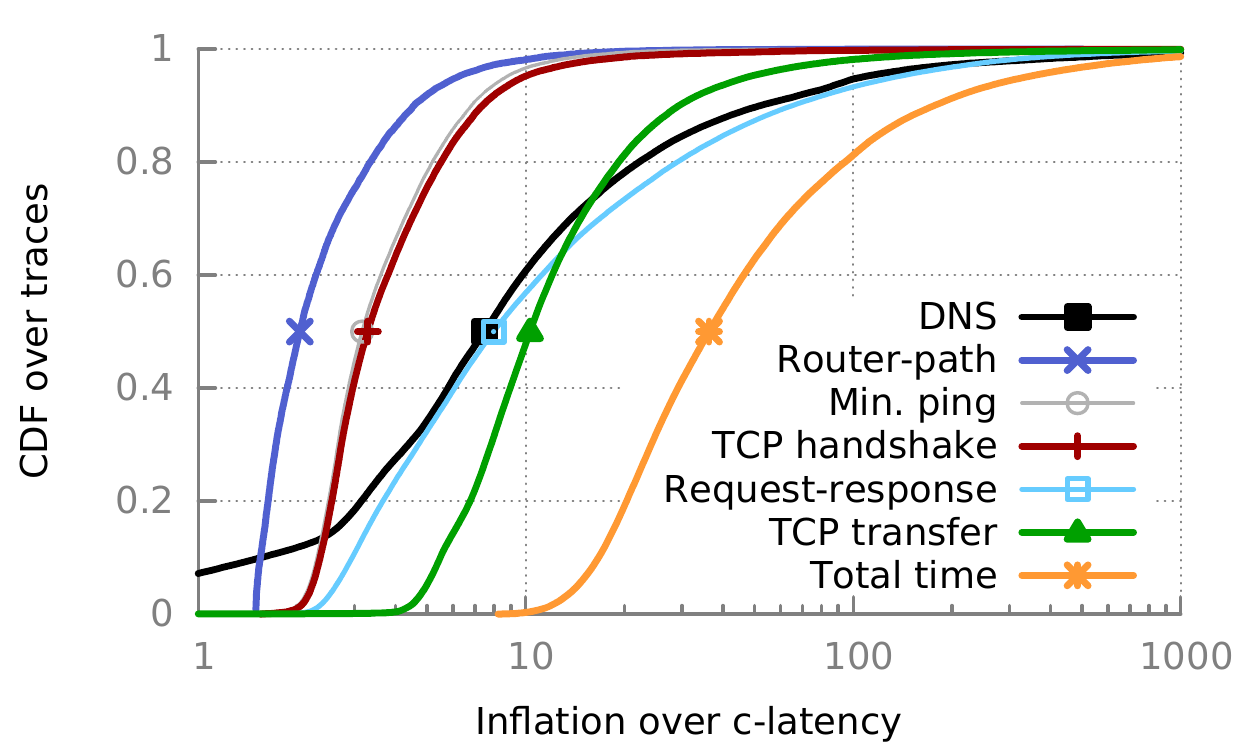}}
\vspace{-8pt}
\caption{\small \em Internet latency inflation measured across page sizes in our measurements: (a) the distribution of Web page sizes in our traces; (b) inflation in minimum ping time, TCP transfer time, and total time as a function of page size; and (c) various components of latency inflation across pages within $10\%$ of the median page size.}
\label{fig:size_inflation}
\vspace{-8pt}
\end{figure*}

It is worth noting that the medians of inflation in DNS time, TCP handshake time, request-response time, and TCP transfer time add up to $27.6\times$, in comparison to the \emph{measured} median total time of $35.4\times$. We should expect such a discrepancy because of the distributions being tail heavy.

\subsection{Impact of IP geolocation errors}
\label{subsec:geolocation}

%Before we dive into an analysis of latency inflation due to congestion (\S\ref{subsec:congestion}), and inflation at the physical and routing layers (\S\ref{subsec:infra}), below we check our results for consistency across possible geolocation errors, the spread of page sizes, and host-pair $c$-latencies. \fixme{structure is confusing}
%\paragraphb{Errors in IP geolocation:}
While we cull data with obvious anomalies arising from geolocation errors (such as when the minimum ping time is smaller than the $c$-latency computed based on IP geolocation), less obvious errors could impact our results. Obtaining ground truth information for the large IP space under consideration appears infeasible. Thus, we focused our effort on comparing the results we obtained by using $5$ different commercial IP geolocation services, as well as a location computed as their majority vote (MV). We computed latency inflation in router-path, minimum ping, and total time using each of these $6$ sets of IP geolocations. Fig~\ref{fig:geolocation} shows the comparison; as we might expect, router-path latency (Fig.~\ref{fig:geo-rp}) is most susceptible to differences in IP geolocation --- the result there depends on geolocating not only the Web server, but also each router along the path. Even so, all $6$ median inflation values are in the $2$-$2.4\times$ range. Differences in results for minimum ping time (Fig.~\ref{fig:geo-minp}) and total time (Fig.~\ref{fig:geo-total}) are much smaller. Even the $95^{th}$-percentile values for inflation in minimum ping time all lie within $10.1$-$11.0\times$, while the medians lie within $3.2$-$3.3\times$. The results for median inflation in total time all lie between $35.4$-$37.6\times$, but variation at the higher percentiles is larger. With the exception of Fig.~\ref{fig:geolocation}, we use the majority vote geolocation throughout.

%Needless to say, we cannot, without ground truth, account for systematic errors that may impact all geolocation services. However, Geo-$5$ does not match the majority vote location as much as $53\%$ of the time, with a generous definition of `match', allowing up to $3$ degrees of difference in both latitude and longitude. Yet, the aggregate results in Fig.~\ref{fig:geolocation} show little difference. Thus, our conclusions, particularly with regards to median values are robust to significant errors in geolocation.

%\paragraphb{Effect of fetch size:}
\subsection{Results across page sizes}
\label{subsec:sizes}
While we only fetch the HTML for the landing pages of Web sites in our experiments, some of these are still larger than $1$MB. However, as Fig.~\ref{fig:fetchsize:dist} shows, most pages are much smaller, with the median being $47$KB. To analyze variations in our results across page sizes, we binned pages into $1$KB buckets, and computed the median inflation for each latency component across each bucket. Fig.~\ref{fig:fetchsize:sizeInfl} shows the median inflation in ping time, TCP transfer time, and total time across different page sizes. The median inflation in minimum ping time shows little variation across page sizes, as one might expect. Inflation in TCP transfer time increases over page sizes in an expected fashion, also causing an increase in total fetch time.

\begin{figure*}
\centering
\subfigure[]{ \label{fig:cspeed:dist}    \includegraphics[width=2.25in]{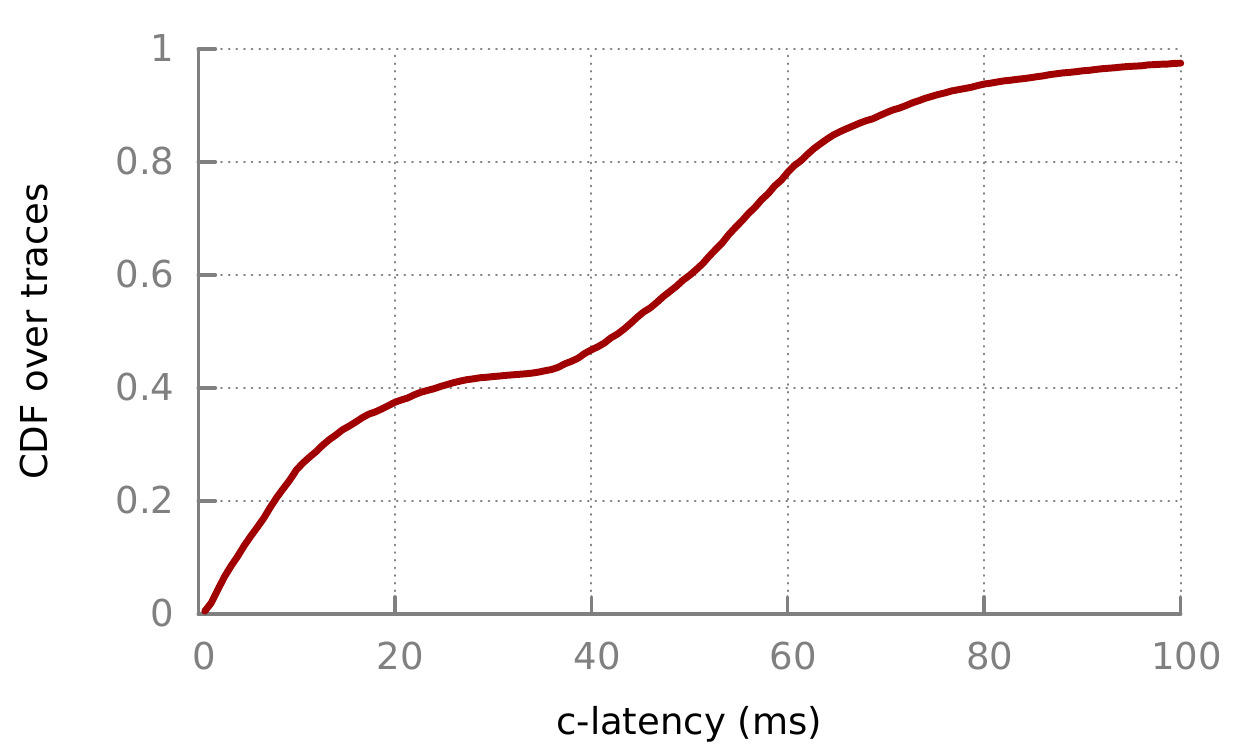}}
%\vspace{-6pt}
\subfigure[]{ \label{fig:cspeed:cspeedInfl}\includegraphics[width=2.25in]{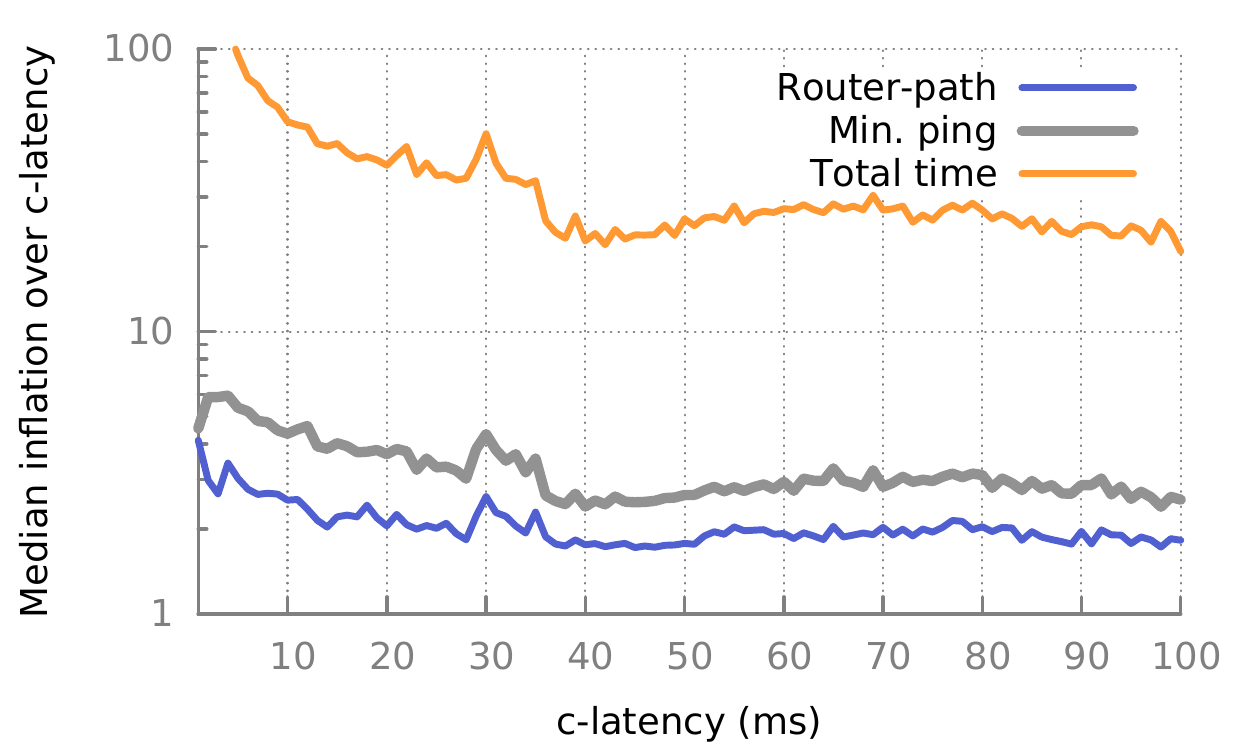}}
%\vspace{-6pt}
\subfigure[]{ \label{fig:cspeed:table} \includegraphics[width=2.25in]{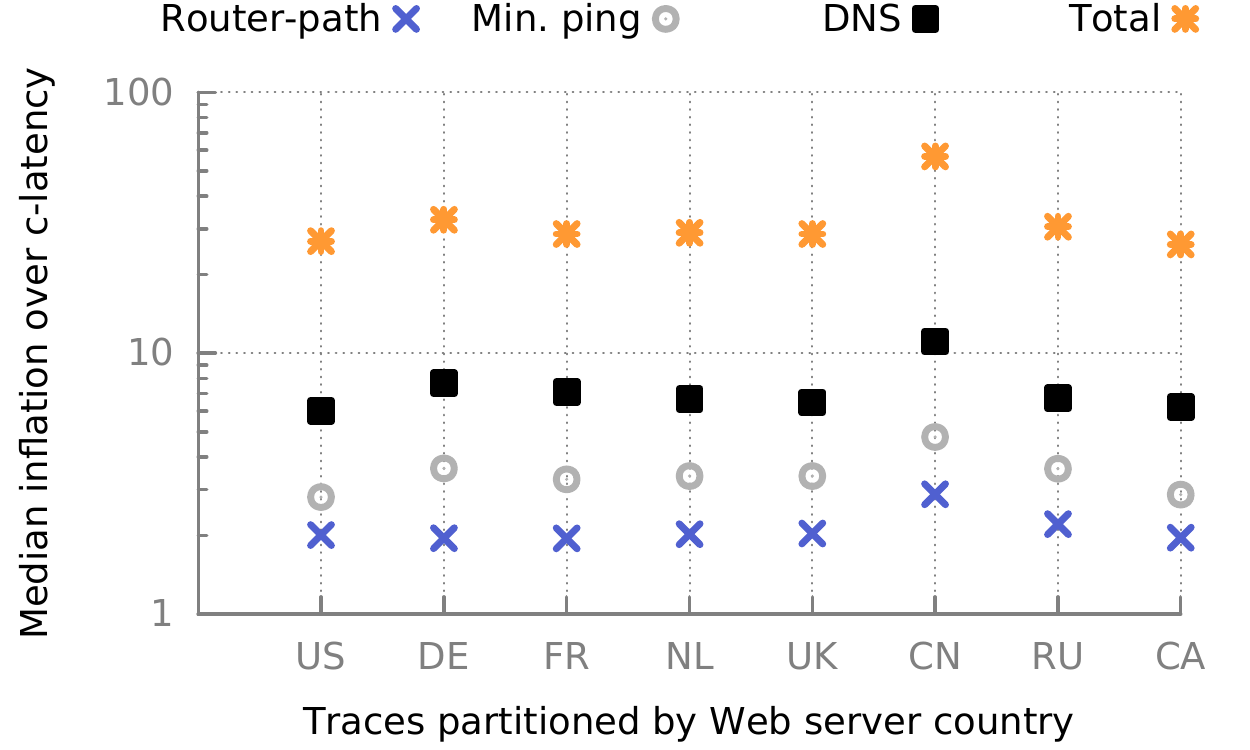}}
\vspace{-8pt}
\caption{\small \em Internet latency inflation measured across $c$-latencies and geographies: (a) the distribution of $c$-latencies across host-pairs in our traces; (b) inflation in router-path latency, minimum ping time, and total time as a function of $c$-latency; and (c) inflation in router-path latency, minimum ping time, DNS time, and total time across $8$ countries.}
\label{fig:cspeed_inflation}
\vspace{-8pt}
\end{figure*}

We also examine latency inflation in a narrow range of Web page sizes around the median, using pages within $10\%$ of the median size of $47$KB. These pages comprise roughly $6\%$ of our dataset. The results of this analysis are shown in Fig.~\ref{fig:fetchsize:medsize}, and are similar to the overall results in Fig.~\ref{fig:prelim}, with expected differences in the transfer time and total time curves. The medians for various components of latency inflation are all within $3\%$ of the results in Fig.~\ref{fig:prelim}, except request-response time where the median is $21\%$ larger for this set.

%\paragraphb{Effect of distance:}

\subsection{Results across geographies}
\label{subsec:geographies}
We fetch pages in $103$ countries from $186$ unique PlanetLab locations, leading to a wide spread in the pairwise $c$-latencies observed across these connections. This variation is captured in Fig.~\ref{fig:cspeed:dist}, which shows the distribution of traces across different $c$-latencies. ($c$-latencies were binned into $1$ms bins for this analysis.) The shape of the curve is largely a result of the Earth's geography and the distribution of our PlanetLab clients. The median $c$-latency is $43$ms. In a manner similar to our analysis across page sizes, we also analyzed latency inflation across $c$-latencies. Fig.~\ref{fig:cspeed:cspeedInfl} shows the median inflation in router-path latency, minimum ping time, and total time. As one might expect, latency inflation is higher for small $c$-latencies. An interesting feature of these results is the inflation bump around a $c$-latency of $30$ms. It turns out that countries such as Portugal, Iran, Ireland, Iceland, and Ecuador, connectivity to which may be more circuitous than average, are over-represented at these distances in our data. For instance, $c$-latencies from the Eastern US to Portugal are in the $30$ms vicinity, but all transatlantic connectivity hits Northern Europe, from where routes may go through the ocean or land Southward to Portugal, thus incurring significant path `stretch'. That the differences are largely due to inflation at the lowest layers is also borne out by the inflation in minimum ping and total time following the inflation in the router-path latency.

%An encouraging observation from Fig.~\ref{fig:cspeed:cspeedInfl} is that the inflation in minimum ping and total time follows the inflation in the router-path latency. Thus, despite the router-path latency estimation containing multiple approximations (omitting routers that did not respond to traceroutes or we could not geolocate, as well as paths between successive routers themselves potentially being circuitous), it is a useful quantity to measure.

For a fairer comparison across geographies, we selected $28$ PlanetLab hosts such that no two were within $5$ degrees of longitude of each other. Then we looked at requests from these PlanetLab clients to Web servers in each country. Fig.~\ref{fig:cspeed:table} shows the median inflation in router-path latency, minimum ping time, DNS, and total time across each of the $8$ countries for which we had $10$,$000+$ connections. The median $c$-latencies (not shown in Fig.~\ref{fig:cspeed:table}) from these selected PlanetLab hosts to each these $8$ countries all lie in the $47$-$54$ms range, with the exception of China ($59$ms). This is likely attributable to the distribution of PlanetLab hosts --- no PlanetLab hosts were available to us between longitudes $38^{\circ}$E-$100^{\circ}$E, thus placing China further away from our hosts on average than other nations. Router-path latencies are fairly consistent across geographies, with the exception of fetches from China, which is also much worse than the others for each of minimum ping time, DNS, and total time, for reasons that are not clear to us\footnote{Ahem ... (Great) ... ahem ... (Firewall)?}. Across the other $7$ countries, median inflation ranges between $2.8$-$3.6\times$ for minimum ping time, $6$-$7.7\times$ for DNS, and $26.1$-$32.5\times$ for total time.

%Having analyzed the somewhat easier to examine TCP and DNS factors, w
%We devote the rest of this section to a closer look at inflation in due to physical infrastructure and routing, and congestion.

\subsection{The role of congestion}
\label{subsec:congestion}

Fig.~\ref{fig:prelim} and Fig.~\ref{fig:fetchsize:medsize} show that TCP transfer time is more than $10\times$ inflated over $c$-latency. It is worth considering whether packet losses or large packet delays and delay variations are to blame for poor TCP performance. 
%Oversized and congested router buffers on the propagation path may exacerbate such conditions -- a situation referred to as \emph{bufferbloat}.

In addition to fetching the HTML for the landing page, for each connection, we also sent $30$ pings from the client to the server's address. We found that variation in ping times is small: the $2^{nd}$-longest ping time is only $1.1\%$ larger than the minimum ping time in the median. While pings (using ICMP) might use queues separate from Web traffic, even the TCP handshake time is only $1.6\%$ larger than the minimum ping time in the median. We also used tcpdump~\cite{tcpdump} at PlanetLab clients to log packet arrival times from the servers, and analyzed the inter-arrival gaps between packets. More than $92\%$ of the connections we made experienced no packet loss (estimated as packets re-ordered by more than $3$ms).

For a closer look at congestion in true end-user environments (as opposed to PlanetLab), we examined RTTs in a sample of
TCP connection handshakes between the servers of a large CDN and clients (end
users) over a $24$-hour time period, passively logged at the CDN.  (Most routes to popular prefixes are unlikely to change at this
time-scale in the Internet~\cite{routestable}.)
%The connections under
%consideration here are physically much shorter, making route changes even more
%unlikely.)
We exclude server-client pairs with minimum latencies of less than $3$ms --- `clients' in this latency range are often proxy servers in a data center or colocation facility rather than our intended end-users.
%Due to the prevalence of proxy servers, often deployed to improve
%the browsing experience on mobile clients (for instance, Opera
%Turbo~\cite{opera-turbo}), we filtered out server-client pairs with minimum
%latencies of less than $3$ms; these clients, with low latencies to servers,
%might actually be the proxy servers in a data center or colocation facility
%often hosting CDN servers at the same or a physically nearby facility.

To evaluate the impact of congestion, we examine our data for both variations across time-of-day (perhaps latencies are, as a whole, significantly larger in peak traffic hours), and within short periods of time for the same server-client pairs (perhaps transient congestion for individual connections is a significant problem).
%Congestion can manifest as large variations in either RTTs of
%measurements repeated at different times of the day, or RTTs of repeat
%measurements over a short time period, between a pair of hosts.  
Thus, we discard server-client pairs that do not have repeat measurements. For ease of analysis over time-of-day, we only look at pairs within the same country. Server locations were provided to us by the CDN, and clients were geolocated using a commercial geolocation service.
We include here results for a few geographies that have a large number of measurements after these restrictions.  We bin all RTT measurements into $12$ $2$-hour periods and produce results aggregated over these bins separately for each country.

\begin{figure*}[bth]
  \centering
  \subfigure[]{ \label{fig:tod-var-p50}
    \includegraphics[width=2.25in]{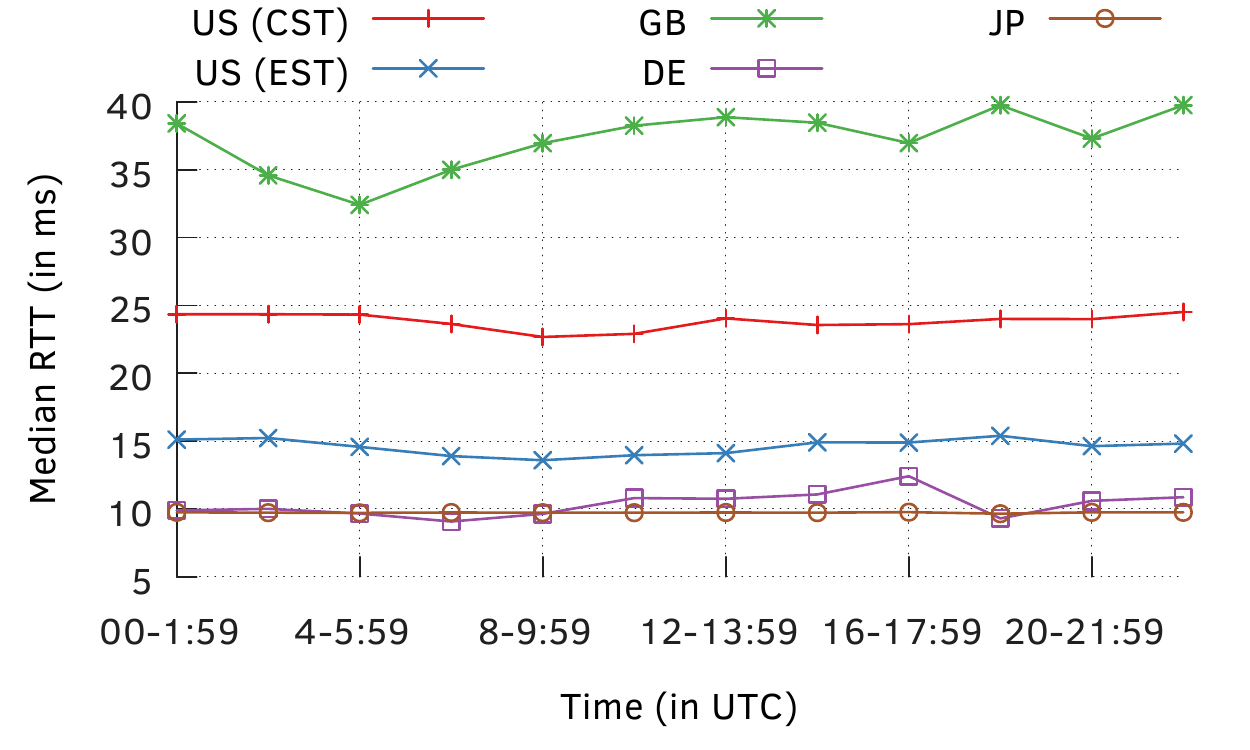}} %\vspace{-6pt}
  \subfigure[]{ \label{fig:tod-var-p90}
    \includegraphics[width=2.25in]{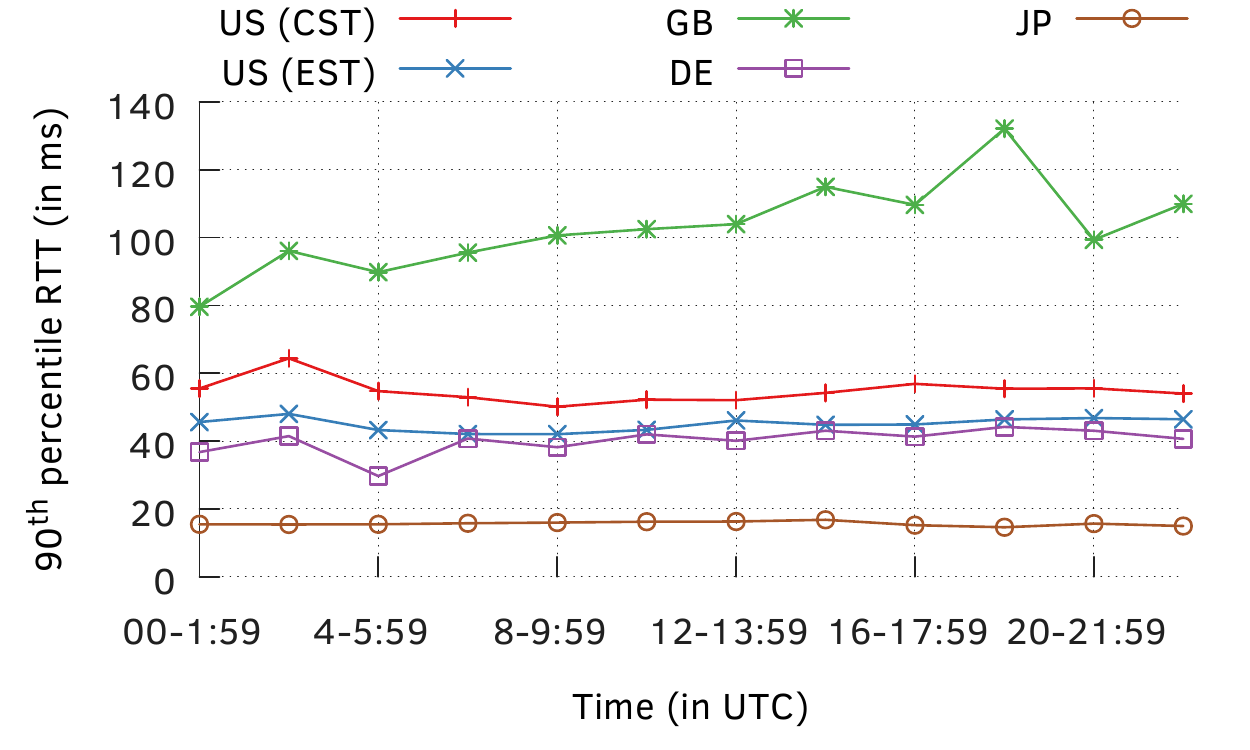}} %\vspace{-6pt}
  \subfigure[]{ \label{fig:dev-var-p50}
    \includegraphics[width=2.25in]{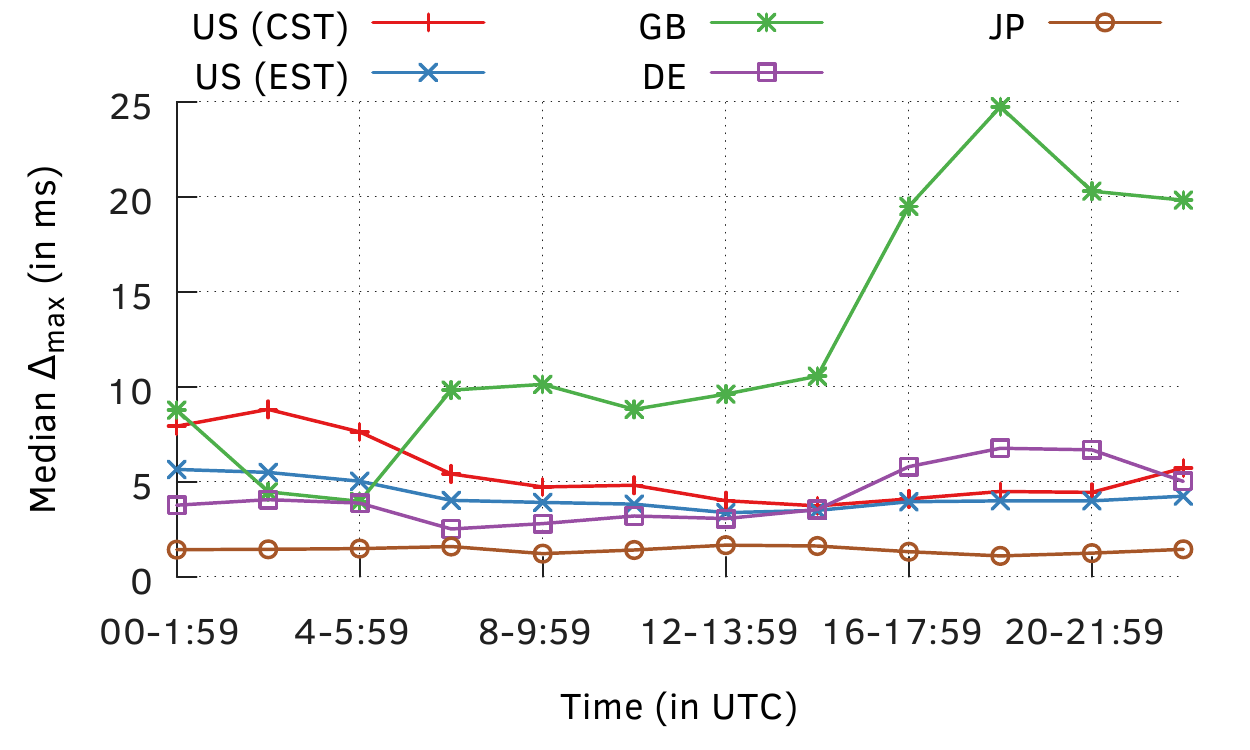}}
\vspace{-8pt}
  \caption{\small \em Variations in latencies of client-server pairs grouped
    into 2-hr windows in different geographic regions: (a) Medians of RTTs of
    client-server pairs with measurements in each 2-hr window; (b) $90^{th}$
    percentiles of RTTs the same set of client-set pairs; and (c) medians of
    maximum change in RTTs (max - min) in repeat measurements within each time
    window.}
  \label{fig:rttvar}
\end{figure*}

\paragraphb{Time-of-day latency variations across bins:} We selected server-client pairs that have at least one RTT measurement in each of the twelve bins.  For pairs with multiple RTTs within a bin, we use the median RTT as representative, discarding other measurements. This leaves us with the same number of samples between the same host-pairs in all bins. Fig.~\ref{fig:tod-var-p50} and Fig.~\ref{fig:tod-var-p90} show the median and the $90^{th}$ percentile of RTTs in each $2$-hour bin for each of $5$ timezones. For the United States (US), we show only data for the central (CST) and eastern (EST) timezones, but the results are similar for the rest\footnote{The timezone
  classification is based on the location of the client; servers associated with
  these measurements can be anywhere in the US and not necessarily restricted to
the same timezone as that of the clients.}. Median latency across our aggregate varies little across the day, most timezones seeing no more than $3$ms of variation, except Great Britain, where the maximum latency difference is $7.35$ms. The $90^{th}$ percentile in each bin (Fig.~\ref{fig:tod-var-p90} shows similar trends, although with larger variations.  Again, in Great Britain, RTTs are higher in the evening. (We checked that results for a different $24$-hour period look similar.) It is thus possible that congestion is in play there, affecting network-wide latencies. However, across other timezones, we see no such effect.

\paragraphb{Latency variations within bins:} To investigate variations within bins, we do not limit ourselves to measurements across the same set of host-pairs across all bins. However, within each bin, only data from host-pairs with multiple measurements inside that time period is included. For each host-pair in each bin, we calculate the maximum change in RTT ($\Delta_{max}$) -- the difference between the maximum and minimum RTT between the host-pair in that time period. We then compute the median $\Delta_{max}$ across host-pairs within each bin. Fig.~\ref{fig:dev-var-p50} shows the results: the variation within bins is a bit larger than variations across median latencies across the day. For example, for US (CST), the median $\Delta_{max}$ is as large as $9$ms in the peak hours. That $\Delta_{max}$ also shows broadly similar time-of-day trends to median latency is not surprising. GB continues to show exceptionally large latency variations, with a $\Delta_{max}$$\simeq$$25$ms at the peak, and also large variations across the day.

To summarize, even in the PlanetLab context, where latency variations and congestion are minimal, Internet latencies show large inflation. In end-user environments, network-wide latency increases in peak hours were largely limited in our measurements to one geography (Great Britain). However, individual flows may occasionally experience a few additional milliseconds of latency.

\subsection{Path inflation}
\label{subsec:infra}

%Fig.~\ref{fig:prelim} shows that the router-path, in the median,
%is only $2.3\times$ inflated. This is small, considering that $1.5\times$ inflation would occur
%occur even along the shortest path along the Earth's surface because the speed of light in fiber is roughly $2/3^{rd}$
%the speed of light in air / vacuum. Again, the tail for the router-path inflation is long. In part,
%this is explained by `hairpinning', \ie packets between nearby
%end-points traversing circuitous routes across the globe. For instance, in some cases, packets between
%end-points in Eastern China and Taiwan were seen in our traces traveling first to California.

Fig.~\ref{fig:prelim} shows that in the median, the router-path is only $2\times$ inflated. The long tail is, in part, explained by `hairpinning', \ie packets between nearby end-points traversing circuitous routes across the globe. For instance, in some cases, packets between end-points in Eastern China and Taiwan were seen in our traces traveling first to California. Note that $1.5\times$ inflation would occur even along the shortest path along the Earth's surface because the speed of light in fiber is roughly $2/3^{rd}$ the speed of light in vacuum. In that light, $2\times$ may appear small. But as we discuss below, our router-path estimate is optimistic, and the lower layers play a significant role in overall inflation.

\begin{figure*}
\centering
\subfigure[]{ \label{fig:fiber:i2}   \includegraphics[width=2.25in]{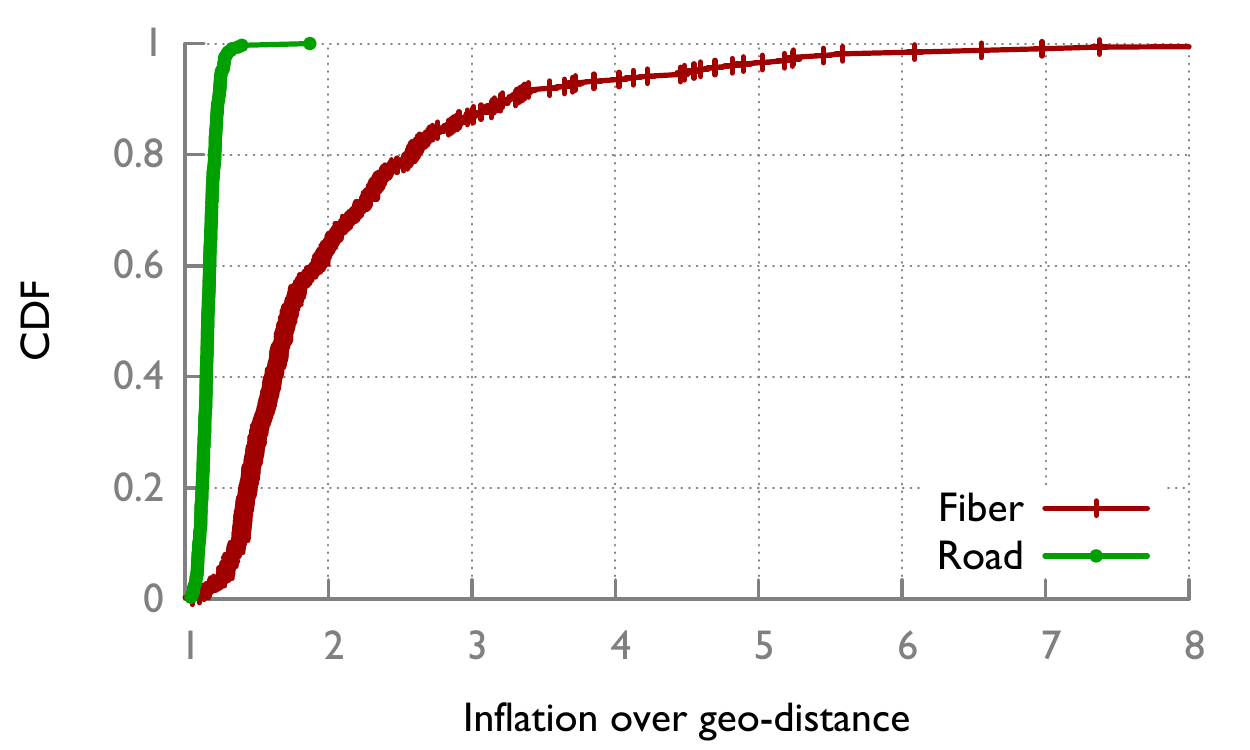}}
%\vspace{-6pt}
\subfigure[]{ \label{fig:fiber:esnet}\includegraphics[width=2.25in]{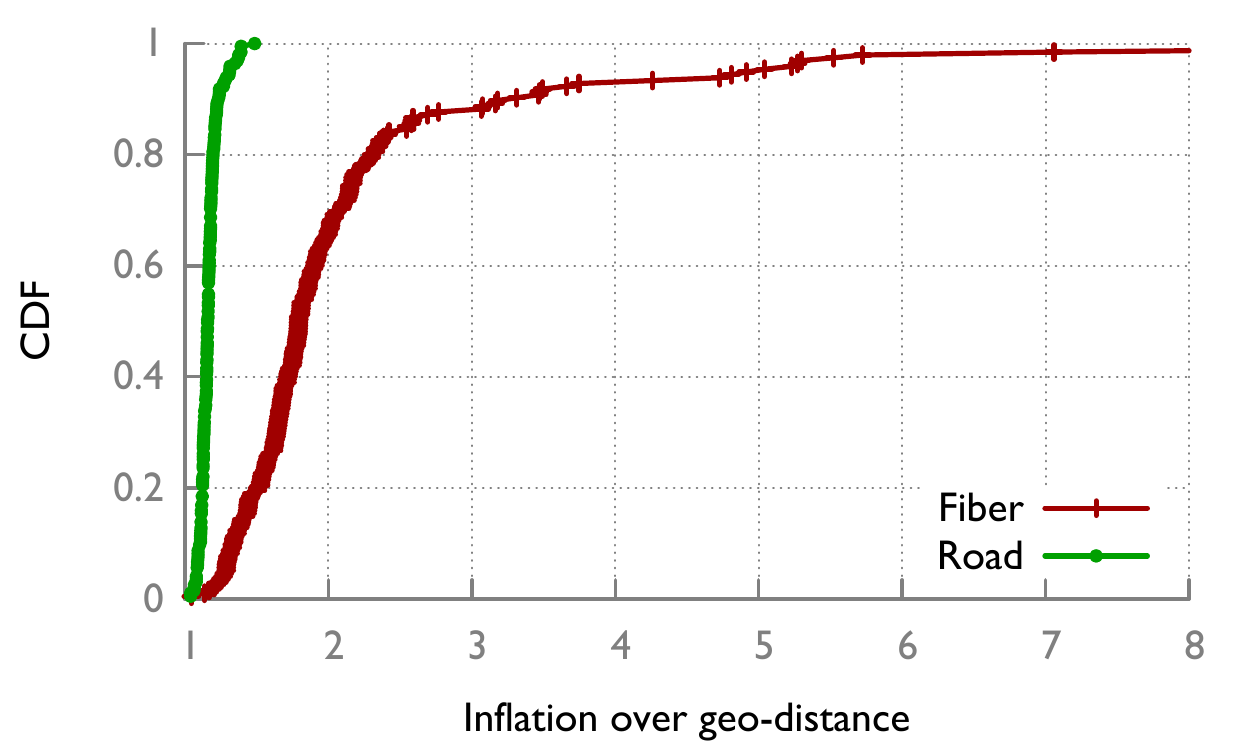}}
%\vspace{-6pt}
\subfigure[]{ \label{fig:fiber:geant}\includegraphics[width=2.25in]{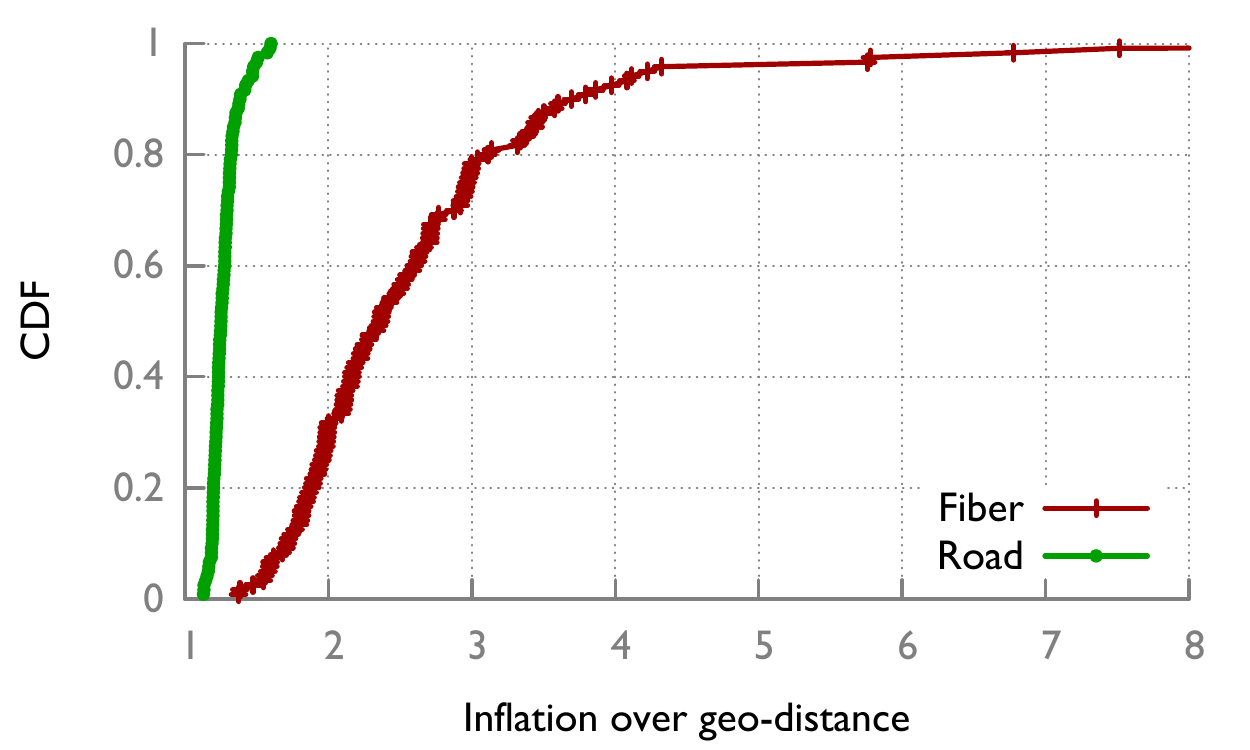}}
\vspace{-8pt}
\caption{\small \em Compared to the shortest distance along the Earth's surface, there is significantly more inflation in fiber lengths than in road distances in all three networks (a) Internet2; (b) ESnet; and (c) G\'{E}ANT.}
\label{fig:fiber}
\vspace{-8pt}
\end{figure*}

We see some separation between the minimum ping time and the router-path latency. This gap may be explained by two factors: (a) traceroute often does not yield responses from all the routers on the path, in which case we essentially see artificially shorter paths --- our computation simply assumes that there is a direct connection between each pair of successive replying routers; and (b) even between successive routers, the physical path may be longer than the shortest arc along the Earth's surface. We investigate the latter aspect using data from three research networks: Internet2~\cite{internet2}, ESnet~\cite{esnet}\footnote{Dhruv Diddi helped process the ESnet data.}, and G\'{E}ANT\footnote{Data on fiber mileages from G\'{E}ANT\cite{geant}, the high-speed pan-European research and education network, was obtained through personal communication with Xavier Martins-Rivas, DANTE. DANTE is the project coordinator and operator of G\'{E}ANT.}. We obtained point-to-point fiber lengths for these networks and ran an all pairs shortest paths computation on each fiber-length annotated network map to calculate end-to-end fiber distances between all pairs of end-points in each network. We also calculated the shortest distance along the Earth's surface between each pair of end-points, and obtained the road distances for comparison using the Google Maps API~\cite{mapsAPI}. Fig.~\ref{fig:fiber} shows the inflation in fiber lengths and road distances compared to the shortest distance. Road distances are close to shortest distances, while fiber lengths are significantly larger and have a long tail.
%Even when only point-to-point connections are considered (i.e. a single fiber link between two cities), fiber lengths are usually $1.5$-$2\times$ larger than road distances.
\ankitnew{The median inflation in the three networks, after accounting for the lower speed of light in fiber (Fig.~\ref{fig:fiber} does not include this adjustment), is $2.6\times$ (Internet2), $2.7\times$ (ESnet), and $3.6\times$ (G\'{E}ANT).
%While research networks may be sparser than commercial ones, even point-to-point fiber-distances, for which there may be little reason to be different across research / commercial settings, are similarly inflated.
Thus, infrastructural inflation (which includes routing sub-optimalities and inflation of end-to-end fiber-distances over geodistance) is likely to be larger than the optimistic estimate from router-path latency ($2\times$), bringing it closer to the inflation in minimum ping latency ($3.2\times$).}
%\fixme{What's the point of these conclusions?  I think we're saying that the ``infrastructure inflation'' (or whatever we want to call it) is very likely closer to the $3.3\times$ min ping time than the $2\times$ median route inflation.  This makes a good transition into the next section.  Also, in addition to (a,b) above, what about (c) congestion?  It's likely not much of the cause because this is the \emph{minimum} ping but we should point that out.}

\paragraphb{Putting inflation at lower layers in perspective:}
%Our results are robust to errors in geo-location, and consistent and meaningful across page fetch sizes and geographies.
As Fig.~\ref{fig:prelim} shows, DNS resolution ($7.4\times$ inflated over $c$-latency), TCP handshake ($3.4\times$), request-response time ($6.6\times$), and TCP transfer ($10.2\times$), all contribute to a total time inflation of $35.4\times$. With these numbers, it may be tempting to dismiss the $3.2\times$ inflation in the median ping time. But this would be incorrect because lower-layer inflation, embodied in RTT, has a \emph{multiplicative} effect on each of DNS, TCP handshake, request-response, and TCP transfer time. The total time for a page fetch can be broken down roughly (ignoring minor factors like the client stack) as:
\begin{multline*}
\vspace{-5pt}
T_{total} = T_{DNS} + T_{handshake} + T_{request} \\
+ T_{server proc} + T_{response} + T_{transfer}
\end{multline*}

If we changed the network's RTTs as a whole by a factor of $x$, everything on the RHS except the server processing time (which can be made quite small in practice) changes by a factor of $x$ (to an approximation; TCP transfer time's dependence on RTTs is a bit more complex), thus changing $T_{total}$ by approximately a factor of $x$ as well.

What if there was no inflation in the lower layers, \ie RTTs were the same as $c$-latencies? For an approximate answer, we can normalize inflation in DNS, TCP handshake, request-response (excluding the server processing time, \ie only the RTT) and TCP transfer time to that in the minimum ping time. Normalized by the median inflation in ping time ($3.2\times$), these numbers are $2.3\times$ (DNS), $1.1\times$ (TCP handshake), $1.0\times$ (request-response, excluding server processing time), and $3.2\times$ (TCP transfer) respectively. The inflation in ping time is at par with the largest of these other (normalized) factors! Further, if, for example, TCP transfer could be optimized such that it happens within an RTT, the Internet would still be worse than $\sim$$25\times$ slower than the $c$-latency in the median, but if we could cut inflation at the lower layers from $3.2\times$ to close to $1\times$, even if we made no transport protocol improvements, we would get to around $\sim$$12\times$. These numbers are, of course, approximate assessments, the larger point being that the contribution of inflation at lower layers is \emph{multiplicative}. Thus, inflation at the lower layers plays a big role in Internet latency inflation, and getting to a speed-of-light Internet requires both infrastructural improvements and protocol advances.

\section{Fast-Forward to the Future}
\label{sec:proposal}

In line with the community's understanding, our measurements affirm that TCP transfer and DNS resolution are important factors causing latency inflation. However, building a speed-of-light Internet requires addressing not only inflation due to protocols, but also that stemming from the Internet's infrastructure. Is this then, a lost cause, as infrastructural problems often are deemed to be? No! In fact, we present here a surprisingly low-cost solution to the infrastructural problem.
%, and argue that contrary to intuition, it might even be easier to address than the protocol factors.

\subsection{A parallel low-latency infrastructure} 
\label{subsec:microwave}

The approach we propose is to build a ``parallel Internet'' to move traffic along nearly the shortest paths on the Earth's surface at nearly the speed of light in vacuum.
How might we build such an alternate Internet infrastructure? In addressing this question, we draw inspiration from the industry which perhaps places the highest premium on milliseconds today: high frequency trading. The HFT industry has already demonstrated the plausibility of operating long-distance links at nearly the speed of light in vacuum. In the quest to cut latency between the New York and Chicago stock exchanges, several iterations of this connection have been built, aimed at successively improving latency by just a few milliseconds at the expense of hundreds of millions of dollars~\cite{chi-ny}. In the mid-$1980$s, the round-trip latency was $14.5$ms. This was cut to $13.1$ms by $2010$ by shortening the physical fiber route. In $2012$ however, the speed of light in fiber was declared \emph{too slow}: microwave communication cut round-trip latency to $9$ms, and later down to $8.5$ms~\cite{fintimes, wiredmag}. The $c$-latency, \ie the round-trip travel time between the same two locations along the shortest path on the Earth's surface at the speed of light in vacuum, is only $0.6$ms less. A similar race is underway along multiple segments in Europe, including London-Frankfurt~\cite{microwaveEurope}.

Today at least $15$ such microwave networks connect the Chicago and New York stock exchanges~\cite{chi-ny}. These networks are built using a series of microwave repeaters mounted on towers along roughly the shortest path along the Earth's surface between the two cities. The `building block' is a microwave repeater mounted on a tower which can span $70$km\footnote{This number depends on terrain, but at least one network in the Chicago-New~York segment uses towers close to this distance from each other~\cite{chi-ny}.} between towers at a data rate of $400$Mbps, with the turnkey installation cost of one repeater ranging from \$$100$,$000$-$250$,$000$ (``including equipment, path engineering, site and path surveys, frequency coordination and licensing, tower work, installation, and commissioning''~\cite{chi-ny}), with an operating cost (including a tower lease, power, maintenance, network operating centers, etc.) of \$$38$,$000$ per year. The typical turnkey installation costs are closer to \$$100$,$000$, but hit the high end estimate in the Chicago-New~York segment due to competition for towers and spectrum.\footnote{Personal correspondence with Gregory Laughlin and Anthony Aguirre, authors of~\cite{chi-ny}.}

\begin{figure*}
\centering
\includegraphics[width=7in]{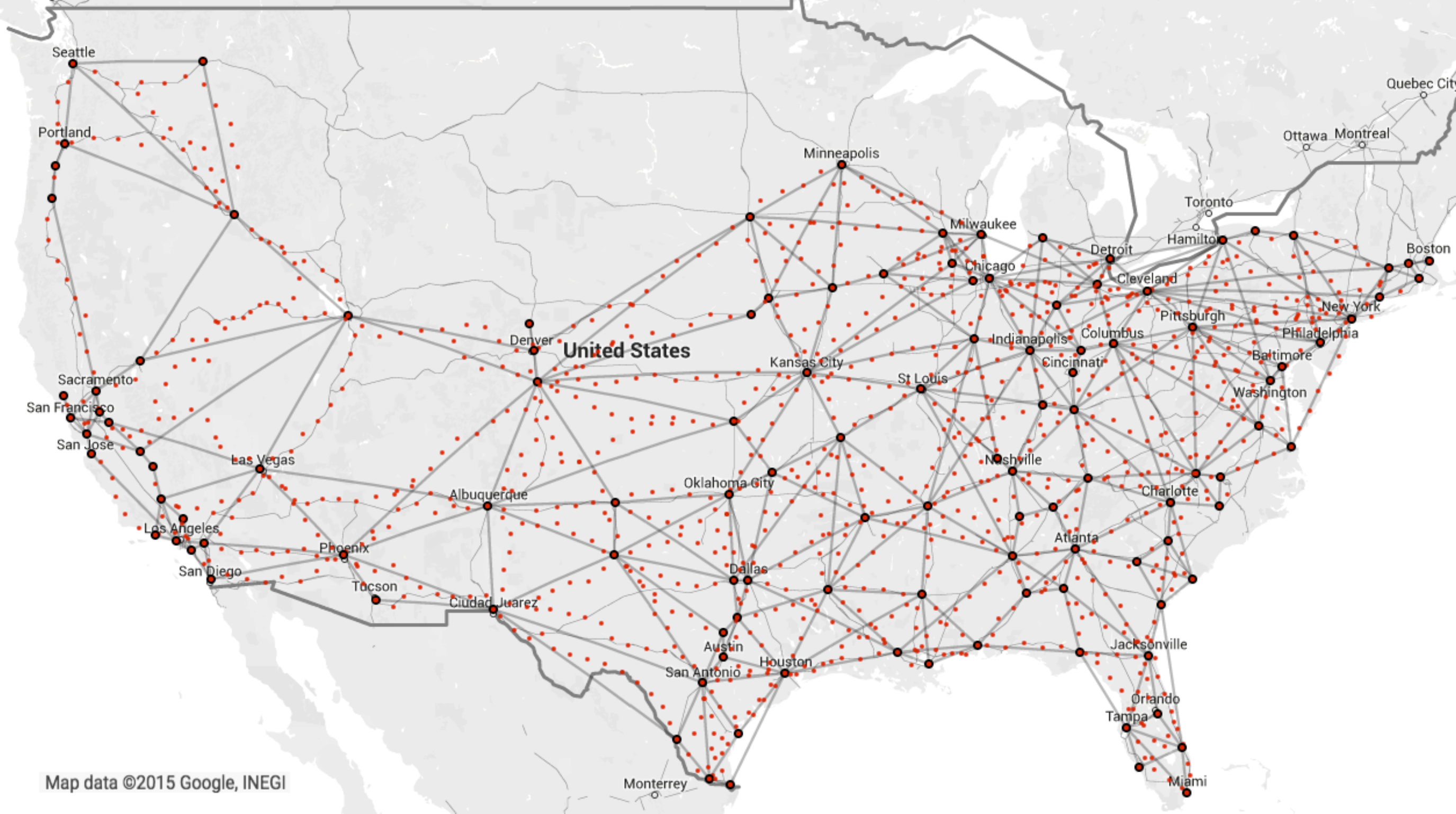}
\vspace{-10pt}
\caption{\small \em A nearly speed-of-light network built across the United States over microwave towers. The gray lines show the logical city-to-city direct microwave connections, which are then realized using the shortest possible chain of microwave towers between each pair of cities connected. Each tower is marked by an individual marker. No two towers on a path are farther than $70$ kilometers.}
\label{fig:networkmap}
\vspace{-8pt}
\end{figure*}

In short, our proposal is to emulate this idea on a wide scale \ankitnew{to build a ``$c$-network''. We could connect pairs of cities using a series of microwave towers. Henceforth, we refer to such a city-to-city microwave connection as an `edge', with a $c$-network being a network of such edges. An existing Chicago-New~York edge, for instance, uses $18$ towers in this manner~\cite{chi-ny}. At its end-point cities, each edge would connect to routing infrastructure, thus becoming part of a network with wide geographic coverage. Of course, fashioning a practical solution from this approach requires careful decisions about the extent of such infrastructure's geographical coverage, and its capacity and usage. In the following, we address these design decisions first in the abstract, and then with a case study of such a proposal limited to the US.}

%The cost for one such end-to-end network connection is estimated at roughly $\$5$ million in initial expenses (including site and path surveys, tower construction / lease, spectrum licensing, etc.) followed by $\$0.7$ million annually for operating expenses. 

\paragraphb{Geographical coverage:} Building a ubiquitous speed-of-light Internet with today's technology is likely to be cost-prohibitive. A more reasonable approach is to connect centers of dense population into a $c$-network, and use traditional connectivity such as fiber to reach areas up to $100$km or so from these centers. In the typical case, the two communicating end-points would use traditional Internet infrastructure to reach their closest $c$-networked centers, from where traffic would be carried over the $c$-network. Using fiber over $100$km incurs an additional $\frac{1}{3}$ms of latency round-trip in comparison with $c$-connectivity. Thus, even if the physical paths in these final $100$km of geo-distance are somewhat circuitous, we can expect the total latency overhead to be limited to $2$-$3$ms for such an end-to-end connection. The goal should be to cover the large distances at nearly $c$. 

Obviously, some scenarios pose difficulties for such a deployment. Building trans-Atlantic microwave connectivity, for example, will require further innovation, although the HFT industry is already considering the possibility of using weather balloons to overcome that challenge~\cite{transAtlantic}. \nnn{Over time, other technologies which operate at nearly the speed of light in vacuum (such as the under-development ``hollow fiber''~\cite{hollowfiber}) may come to fruition, but microwave appears presently to be the only mature, cheap solution.}

\paragraphb{Capacity and usage:} Matching the present Internet's bandwidth in a parallel speed-of-light infrastructure would be exorbitantly expensive, perhaps even impossible --- there might not even be enough wireless spectrum to be able to accomplish this with today's technology. However, most traffic on the Internet is not latency sensitive. For example, video streaming, file sharing, and software downloads comprise an overwhelming fraction of the traffic on the Internet. Further, even most interactive Web traffic consists of small requests (few KB) which fetch responses larger by two to three orders of magnitude. While small responses may be accommodated over the $c$-network, larger responses could be sent over traditional connectivity. If protocol overheads were negligible, this would still achieve $\frac{c}{2}$-connectivity overall. Thus, targeting $1\%$ of the latency sensitive Web traffic's capacity needs would be a reasonable goal. 

\nnn{Admittedly, some latency-sensitive applications we discuss in \S\ref{sec:nfs} (such as tele-immersion) depend on both high bandwidth and low latency. While our present proposal may be inadequate for such applications at scale due to limited bandwidth, it still serves other applications such as instant access. We consider this work a start in this direction.}
%Our $c$-infrastructure proposal should thus target a small fraction of today's Internet's bandwidth needs, limiting its use to latency sensitive applications alone.

In some sense, this infrastructure may be easier to make progress on than many protocol changes, which depend on consensus among several stakeholders. It also has a healthy incremental deployment potential. An ISP might start by selling a low-latency transit service to other ISPs on a few critical routes. It might also be possible to market directly to end-users by making use of tunneling in a manner suggested by a recent research proposal~\cite{onetunnel}, where a user sends their packets first to an IP address connected to the $c$-network, which tunnels them to the exit from the $c$-network which is nearest to the packets' final destination. \nnn{(In this sense, the $c$-network uses ``cold potato'' routing.)}

In the following section, we analyze the coverage, capacity, and cost of a nearly speed-of-light network across the contiguous United States.

\subsection{A $c$-ISP in the US}

\paragraphb{Geographical coverage:} We focus on connecting only the $200$ most populous cities in the contiguous United States
%\footnote{In an innocuous instance of experimenter bias, Champaign-Urbana was included at the expense of Topeka, Kansas at the bottom of our list.} 
with each other over this infrastructure. Further, we coalesce suburbs and cities in close vicinity of each other (within $50$ kilometers), ending up with $120$ population centers. Based on population data for $2010$~\cite{populationData}, we calculate that $\sim$$258$ million people comprising $\sim$$85\%$ of the US population live within $100$ kilometers of these $120$ population centers. 

\paragraphb{Capacity and usage:} Cisco's forecast for ``Consumer Internet Traffic'' for $2015$ in North America\footnote{An equivalent number for the contiguous US was not available. We note that the US population is only two-thirds of that of North America, but continue to use Cisco's North America traffic estimate without any proportional adjustment as a conservative upper bound for our analysis.} is $\sim$$11$,$500$ PB/month, \ie $\sim$$4.5$TBps~\cite{ciscoForecast}. Cisco also estimates that $78.5\%$ of Global IP traffic is file sharing and video, the rest ($22.5\%$) being classified as ``Web / Data'' traffic\footnote{It is worth noting that traffic such as software downloads (except when over P2P software) is included in the this category, but is not latency sensitive.}. This leaves a ceiling of $1$TBps for interactive Web traffic, $1\%$ of which is $80$Gbps. Cisco's report also claims that Web traffic is ``spread evenly throughout the day'', unlike video, which has a ``prime time'' in each geography. Thus, we target $80$Gbps as our $c$-ISP's capacity. 

%\paragraphb{Minor latency inflation:} Connecting even $120$ large population centers to each other at the speed of light would require nearly all pairs of connections between them, which would be prohibitively expensive. However, as we shall see later, if we accept fairly minor latency inflation, we can build the $c$-infrastructure at a much more reasonable cost.

%\vspace{6pt}
%\noindent\emph{How then do we connect the largest $120$ population centers in the contiguous US at nearly the speed of light in vacuum?}
%\vspace{2pt}

In building the $c$-ISP, our objective is to ensure that end-to-end paths between all pairs of locations are as close to the shortest paths along the Earth's surface as possible (\ie \emph{path stretch} close to $1$). Our budget is limited largely by the total number of microwave towers needed. 
%We will return to a discussion of network capacity later, focusing for the interim on stretch and the tower count.

While, tower leasing companies advertise a willingness to build towers for long-term leases, we restrict ourselves to using towers that are registered in FCC's `Antenna Structure Registration' database~\cite{fccasr}\footnote{Note that FCC only requires registrations for towers about $200$ feet, or otherwise, in sensitive locations such as near airports. Enforcement of registrations is also difficult, so there are certainly more towers suitable for antennas than in this database.}, with a status of `Constructed'. While many of these towers may not be available for leasing, their existence, at the least, indicates the suitability of their near vicinity for the construction of other towers. We thus focus on using these towers as the foundation of our US-wide microwave network.

We used an intuitive heuristic to design the network, starting with a minimum cost spanning tree (with the cost being number of towers needed to connect a pair of cities), and repeatedly augmenting this network with the edge that minimizes the $95^{th}$ percentile stretch across all bytes that would move through the network. In the traffic matrix we used, traffic between each pair of cities is proportional to the product of their populations. We normalized the traffic matrix such that the total traffic hit our $80$Gbps capacity requirement. Of course, some edges end up with more load than their $400$Mbps capacity --- we replicate each edge enough times not only to accommodate its load, but also to operate at $50\%$ utilization to allow for traffic variations. (Replicating an edge once implies using a parallel series of towers along the route, thus doubling the tower-count.) Our designed network connects $300$ city-pairs with microwave edges built over a total of $2526$ towers (after accounting for any necessary replications), and is shown in Fig.~\ref{fig:networkmap}. 

\begin{figure}
\centering
\includegraphics[width=2.6in]{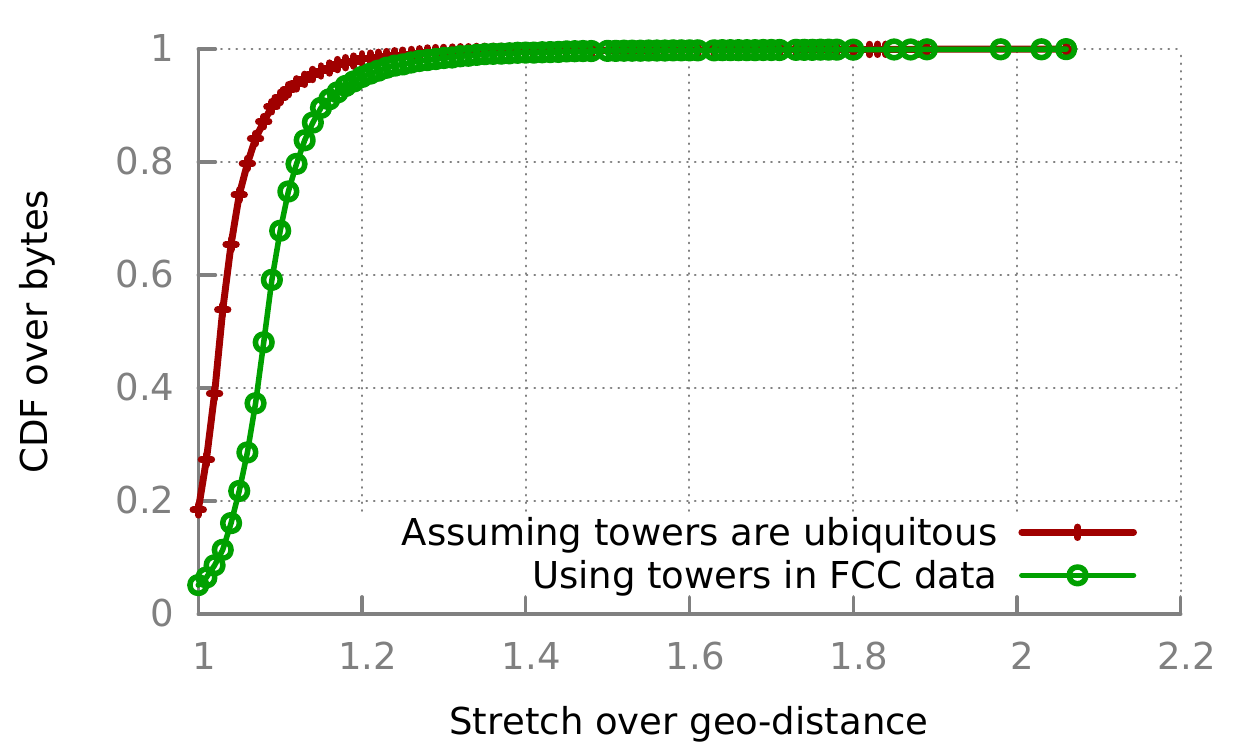}
\vspace{-8pt}
\caption{\small \em The path stretch in our designed network is small.}
\label{fig:proposal_stretch}
\end{figure}

Fig,~\ref{fig:proposal_stretch} shows the path stretch incurred by bytes across the entire network. We show the stretch results for two scenarios: one where towers are ubiquitous, \ie edges between cities are along the geodesics; and another using only the towers in the FCC database. The median and $90^{th}$-percentile stretch numbers are $1.03\times$ and $1.09\times$ (with towers being ubiquitous) and $1.08\times$ and $1.15\times$ (with only the towers in FCC's data). Thus, most bytes are carried at close to $c$-latency. It is also worth noting that our stretch results are not too sensitive to the $70$km range of a microwave repeater, although, obviously, the number of towers required increases as the range decreases. Limiting the range to $50$km results in a network with $3320$ towers and median and $90^{th}$-percentile stretch values of $1.1\times$ and $1.18\times$, while at $40$km, we need $4296$ towers for stretch values of $1.12\times$ and $1.41\times$. When the range is limited to $30$km, some cities are disconnected from the network.

At a \$$100$,$000$ installation cost per tower, this network would cost \$$253$ million, with a \$$96$ million per year operating cost. Amortized over $5$ years, the yearly cost would be $\sim$\$$147$ million. In comparison, the Internet2 predecessor, Abilene, cost \$$500$ million ($1998$, unadjusted)~\cite{abileneCost}. That other ambitious projects with substantially larger price-tags are making progress is also encouraging; among these are OneWeb's plan to expand the reach of the Internet using a total of $648$ satellites in low-Earth orbit at an estimated cost of \$$1.5$-$2$ billion~\cite{arsGlobalInternet}, and a broadly similar effort from WorldVu, cost estimates for which range from \$$10$ billion to $15$ billion~\cite{arsGlobalInternet, muskTelegraph}. Arctic Fiber is spending \$$850$ million to build a submarine cable through the Arctic, focused primarily on shortening the London-Tokyo route~\cite{arcticCable}.

Obviously, our analysis here is coarse, with many opportunities for refinement of the network's design. However, the larger point is that a nearly speed-of-light infrastructure is feasible to build today relatively cheaply.
%\footnote{For context, it is worth reminding oneself that Instagram is valued at several billion USD! A $c$-ISP has much greater utility, besides also being much more `instant'!}.

It is worth mentioning that microwave is not the most reliable medium in inclement weather. However, in the Chicago-New~York segment, which is not particularly known for clear weather, the operational networks claim a $95\%$ uptime, with even low estimates of availability being over $85\%$~\cite{availabMicro}. Further, we continue to have backup connectivity over fiber, and even if a microwave edge connecting two cities is unavailable, the latency gains from the rest of the infrastructure continue to provide value. 

\subsection{A one RTT transaction protocol}
\label{subsec:protocols}

As we note in \S\ref{sec:whyslow}, while eliminating infrastructural latency inflation will speed up the Internet by more than $3\times$, without any progress on transport protocols, this may still leave us more than $10\times$ away from $c$-latency in the typical case. Thus, while the main contributions of this work are our measurements and analysis highlighting latency inflation due to infrastructural inefficiencies, and our proposal for a nearly speed-of-light physical infrastructure, we also discuss here the possibility of building a one RTT transaction protocol for the Internet by putting together various protocol advancements the community's research has produced.

\paragraphb{DNS:} One possibility to speed up DNS lookups is simply brute replication of DNS infrastructure. Along the lines we argued in \S\ref{sec:nfs}, the $c$-infrastructure helps reduce the number of replicas needed to achieve a latency bound -- for example, within the contiguous US, just $7$ replicas would be enough to serve most of the population within $5$ms. \cut{Further, given our observations about the impact of tail DNS latencies, work that suggests making multiple redundant DNS queries from different DNS providers and using the fastest result~\cite{vulimiri} seems promising.} An alternate approach is to try achieving `on-path' lookups in a manner proposed by ASAP~\cite{asap}, where the client sends its data request itself to the DNS server, which resolves the destination's address and forwards the data request on behalf of the client. Another method of achieving `on-path' lookups is name-based routing. However these latter methods have their limitations, requiring extensive architectural changes.

\paragraphb{Eliminating the TCP handshake:} In the absence of a handshake, a malicious client may craft packets with a source address different from its own, thus causing the server to send the response to that address. This technique can be used to attack both the server and other end hosts (which may receive large volumes of data from several servers directed to them by malicious clients). The handshake, by requiring that the client acknowledge a message with a nonce before any significant compute or network resources are spent by the server, prevents this. Avoiding the handshake thus entails address authentication by some other means. Several solutions are available, including ASAP~\cite{asap}, TCP Fast Open's Cookie mechanism~\cite{fastopen}, and APIP~\cite{apip}, with the broad idea being that the server (or another accountable third-party) validates the client's claim to the address (or, in APIP's case, the data packets); subsequently, this validation is included with future requests thus eliminating the handshake for those requests. The latency cost of validation may be incurred much less frequently than with handshaking for each connection.

\paragraphb{Eliminating TCP slow-start:} As we see in \S\ref{sec:whyslow}, TCP's slow start mechanism implies that fetching even just a few tens of kilobytes requires several RTTs. A variety of possibilities exist for handing this problem, the simplest proposal being to use larger TCP initial window sizes~\cite{dukkipati}. In our scenario, one could consider sending a larger TCP window's worth of data over the $c$-infrastructure, and another few tens of kilobytes (if necessary) across traditional connectivity. There are of course, other options, such as Jumpstart~\cite{jumpstart}, which allows for any `appropriate' rate from the beginning, where the appropriate rate may be determined in several possible ways; for example, ISPs could maintain a per-prefix listing of suitable window sizes based on recent history, etc. Another proposal addressing the slow-start problem, RC$3$~\cite{rc3} suggests using multiple priority levels, sending out a small volume at the highest priority, and an exponentially increasing volume of data at successively lower priority levels, thus sending out a large volume of data right from the start, while removing the possibility of congestion collapse. (In RC$3$, packets at each priority obey TCP's congestion collapse directives.) RC$3$ does however pose deployment hurdles, requiring that all routers support multiple priorities.

\paragraphb{TLS/SSL:} In addition to the various protocol pieces above, making \emph{secure} connections quickly is also an important problem. Google's experimental QUIC protocol~\cite{quic} may present a solution here. QUIC proposes that once a client has a server's public key (perhaps through a prior contact, which does require an additional round-trip), it can send its data request, together with a proposed session key, all encrypted with the server's public key. The server can accept the session key (in the common case) and encrypt the response with it. Of course, QUIC too requires the elimination of the handshake and slow-start, and how exactly each objective will be met is currently being worked out. 
%We believe that obtaining the server's public key can be rolled into the DNS resolution step, where the DNS server responds to the client with both the IP address and public key of the server.

QUIC's modular design makes it a useful vehicle for protocol improvements such as eliminating the TCP handshake and folding in a more aggressive congestion control algorithm. Encouragingly, Google's servers already support it, and while it is at this stage experimental, with several pieces of the design yet to be frozen, clients using Google's Chrome browser or Opera can enable it~\cite{quic}.

\nnn{The $c$-network could also spur the development and use of new protocols, for instance, by using new protocols internally and deploying proxies at the edge. This would be similar to what Akamai's SureRoute~\cite{sureRoute} does (in terms of maintaining persistent TCP connections between Akamai servers.)}

\vspace{-10pt}
%\vspace{-6pt}
\section{Related Work}

There is a large body of work on reducing Internet latency. However, this work has been limited in its scope, its scale, and most crucially, its ambition. Several efforts have focused on particular pieces; for example,~\cite{fastopen,asap} focus on TCP handshakes;~\cite{dukkipati} on TCP's initial congestion window;~\cite{vulimiri} on DNS resolution;~\cite{wolfgang, lixingao} on routing inflation due to BGP policy. Other work has discussed results from small scale experiments; for example,~\cite{sundaresan} presents performance measurements for $9$ popular Web sites;~\cite{habib} presents DNS and TCP measurements for the most popular $100$ Web sites. The WProf~\cite{wprof} project breaks down Web page load time for $350$ Web pages into computational aspects of page rendering, as well as DNS and TCP handshake times. Wang et al.~\cite{wang} investigate latency on mobile browsers, but focus on the compute aspects rather than networking.

The central question we have not seen answered, or even posed before, is \emph{`Why are we so far from the speed of light?'}. Even the ramifications of a speed-of-light Internet have not been explored in any depth --- how would such an advance change computing and its role in our lives? Answering these questions, and thereby helping to set the agenda for networking research in this direction is one of our primary objectives. 

The $2013$ Workshop on Reducing Internet Latency~\cite{workshopRedLat} focused on potential mitigation techniques, with bufferbloat and active queue management being among the centerpieces. One interesting outcome of the workshop was a qualitative chart of latency reduction techniques, and their potential impact and feasibility (Fig.~$1$ in ~\cite{workshopRedLat}). In a similar vein, one objective of our work is to \emph{quantify} the latency gaps, separating out factors which are fundamental (like the $c$-bound) from those we might hope to improve.  The goal of achieving latencies imperceptible to humans was also articulated~\cite{tahtLatency}. We share that vision, and in \S\ref{sec:nfs} discuss the possible impacts of that technological leap.

Further, beyond staking out the problem and discussing the potential impacts of a speed-of-light Internet, our measurements and analysis put the focus on an aspect of the latency problem that has been largely ignored so far: infrastructural inefficiencies. We have not seen any work from the community directed at tackling the infrastructural inefficiencies that contribute to latency inflation. There are other ambitious projects related to enhancing Internet infrastructure, like the satellite Internet efforts of OneWeb and WorldVu~\cite{muskTelegraph, arsGlobalInternet}, Google's Loon project~\cite{googleLoon}, and Facebook's drones~\cite{fbDrone}, but these are all addressing a different (albeit important) problem --- expanding the Internet's reach to under-served populations. There are also efforts geared at improving bandwidth in existing Internet markets, such as Google Fiber~\cite{googleFiber}. But infrastructural latency has so far only garnered attention in niche scenarios, such as the financial markets, and isolated submarine cable projects aimed at shortening specific routes~\cite{arcticCable,seasCable}. We believe this to be the first proposal to tackle the role of infrastructure in latency at such a wide scale. In fact, we make the case here that infrastructural latency inflation is not even well \emph{understood}, let alone effectively addressed. Our work also shows it to be a surprisingly more tractable problem than one might believe.

\ankitnew{A recent workshop paper~\cite{singla14hotnets} also addressed the necessity for a speed-of-light Internet, and some basic analysis of the causes of latency inflation. This work sharpens the focus on infrastructural latency inflation, putting it in perspective, and proposes a solution to the problem.}

\vspace{-6pt}
\section{Conclusion}

Speed-of-light Internet connectivity would be a technological leap with phenomenal consequences, including the potential for new applications, instant response, and radical changes in the interactions between people and computing. To shed light on what's keeping us from this vision, in this work, we quantify the latency gaps introduced by the Internet's physical infrastructure and its network protocols, and find that infrastructural gaps make as significant a contribution to latency inflation as protocol overheads. Further, we propose a surprisingly cost-effective method of building a nearly speed-of-light Internet infrastructure.

\section*{Acknowledgments} 

We gratefully acknowledge the support of National Science Foundation Awards 1149895 and 1345284, a Google Research Award, and USAF Award FA8750-14-2-0150. We are also thankful to Inder Monga, ESNet; Xavier Martins-Rivas, DANTE (which operates G\'{E}ANT fiber); and John Hicks, Internet2, for making available data on fiber distances. Dhruv Diddi helped us process this data. Our conversations with Gregory Laughlin and Anthony Aguirre were extremely helpful in understanding existing microwave-based networks.

{\bibliographystyle{abbrv}%\vspace{-6pt}
%\setlength{\bibsep}{2pt}
%\footnotesize{
\bibliography{paper}

\begin{thebibliography}{10}

\bibitem{curl}
{cURL}.
\newblock \url{http://curl.haxx.se/}.

\bibitem{esnet}
{ESnet dark fiber data, fiber distances only}.
\newblock Personal correspondence with Inder Monga, ESNet.

\bibitem{geant}
{G\'{E}ANT}.
\newblock \url{http://www.geant.net/}.

\bibitem{googleFiber}
{Google Fiber}.
\newblock \url{https://fiber.google.com/}.

\bibitem{googleLoon}
{Google Loon}.
\newblock \url{http://www.google.com/loon/}.

\bibitem{mapsAPI}
{Google Maps API}.
\newblock \url{http://goo.gl/I4ypU}.

\bibitem{internet2}
{Internet2}.
\newblock \url{http://www.internet2.edu/}.

\bibitem{quic}
{QUIC, a multiplexed stream transport over UDP}.
\newblock \url{http://www.chromium.org/quic}.

\bibitem{microwaveEurope}
{Quincy Extreme Data service}.
\newblock \url{http://goo.gl/wSRzjX}.

\bibitem{tcpdump}
{tcpdump}.
\newblock \url{http://www.tcpdump.org/}.

\bibitem{visualThresh}
{Temporal Consciousness, Stanford Encyclopedia of Philosophy}.
\newblock \url{http://goo.gl/UKQwy7}.

\bibitem{msMatter}
{The New York Times quoting Microsoft's ``Speed Specialist'', Harry Shum}.
\newblock \url{http://goo.gl/G5Ls0O}.

\bibitem{alexaByCountry}
{Top 500 Sites in Each Country or Territory, Alexa}.
\newblock \url{http://goo.gl/R8HuN6}.

\bibitem{workshopRedLat}
{Workshop on Reducing Internet Latency, 2013}.
\newblock \url{http://goo.gl/kQpBCt}.

\bibitem{abiIOT}
{ABI Research}.
\newblock More than 30 billion devices will wirelessly connect to the internet
  of everything in 2020.
\newblock \url{http://goo.gl/Lnly5a}.

\bibitem{wiredmag}
J.~Adler.
\newblock {Raging Bulls: How Wall Street Got Addicted to Light-Speed Trading}.
\newblock \url{http://goo.gl/Y9kXeS}.

\bibitem{sureRoute}
{Akamai}.
\newblock {Accelerating Dynamic Content with Akamai SureRoute}.
\newblock \url{http://goo.gl/bUh1s7}.

\bibitem{akamaiState}
Akamai.
\newblock {Akamai's State of the Internet}.
\newblock \url{http://goo.gl/ntyy33}, 2014.

\bibitem{muskTelegraph}
N.~Allen.
\newblock {Elon Musk announces `space Internet' plan}.
\newblock \url{http://goo.gl/uEFG8K}.

\bibitem{brutlag}
J.~Brutlag.
\newblock {Speed Matters for Google Web Search}.
\newblock \url{http://goo.gl/t7qGN8}, 2009.

\bibitem{availabMicro}
{BusinessWire}.
\newblock {Anova Technologies founder and CEO discusses wireless connectivity
  in the new book ``Architects of Electronic Trading"}.
\newblock \url{http://goo.gl/wS1rR2}.

\bibitem{pewsurveyGigabit}
P.~R. Center.
\newblock Killer apps in the gigabit age.
\newblock \url{http://goo.gl/NJlMUx}, 2014.

\bibitem{populationData}
{Center for International Earth Science Information Network (CIESIN), Columbia
  University; United Nations Food and Agriculture Programme (FAO); and Centro
  Internacional de Agricultura Tropical (CIAT)}.
\newblock {Gridded Population of the World: Future Estimates (GPWFE)}.
\newblock \url{http://sedac.ciesin.columbia.edu/gpw}, 2005.
\newblock Accessed: 2014-01-12.

\bibitem{chewMusic}
E.~Chew, R.~Zimmermann, A.~Sawchuk, C.~Papadopoulos, C.~Kyriakakis, C.~Tanoue,
  D.~Desai, M.~Pawar, R.~Sinha, and W.~Meyer.
\newblock A second report on the user experiments in the distributed immersive
  performance project.
\newblock In {\em Proceedings of the 5th Open Workshop of MUSICNETWORK:
  Integration of Music in Multimedia Applications}, 2005.

\bibitem{ciscoForecast}
Cisco.
\newblock {Cisco Visual Networking Index: Forecast and Methodology}.
\newblock \url{http://goo.gl/ockPWc}.

\bibitem{abileneCost}
J.~Clausing.
\newblock {New Data Pipeline Holds Promise of a Better Internet}.
\newblock \url{http://goo.gl/uFHnk6}.

\bibitem{fintimes}
C.~Cookson.
\newblock {Time is Money When it Comes to Microwaves}.
\newblock \url{http://goo.gl/PspDwl}.

\bibitem{cuervo}
E.~Cuervo.
\newblock {\em Enhancing Mobile Devices through Code Offload}.
\newblock PhD thesis, Duke University, 2012.

\bibitem{hollowfiber}
{DARPA}.
\newblock {Novel Hollow-Core Optical Fiber to Enable High-Power Military
  Sensors}.
\newblock \url{http://goo.gl/GPdb0g}.

\bibitem{dukkipati}
N.~Dukkipati, T.~Refice, Y.~Cheng, J.~Chu, T.~Herbert, A.~Agarwal, A.~Jain, and
  N.~Sutin.
\newblock {An Argument for Increasing TCP's Initial Congestion Window}.
\newblock {\em {SIGCOMM CCR}}, 2010.

\bibitem{schurman}
{Eric Schurman (Bing) and Jake Brutlag (Google)}.
\newblock {Performance Related Changes and their User Impact}.
\newblock \url{http://goo.gl/hAUENq}.

\bibitem{googVMware}
{Erik Frieberg, VMWare}.
\newblock {Google and VMware Double Down on Desktop as a Service}.
\newblock \url{http://goo.gl/5quMU7}.

\bibitem{fccasr}
{Federal Communications Commission}.
\newblock {Antenna Structure Registration Database}.
\newblock \url{http://goo.gl/3OIFDT}.

\bibitem{lixingao}
L.~Gao and F.~Wang.
\newblock {The Extent of AS Path Inflation by Routing Policies}.
\newblock {\em {GLOBECOM}}, 2002.

\bibitem{gartnerIOT}
Gartner.
\newblock Gartner says the internet of things installed base will grow to 26
  billion units by 2020.
\newblock \url{http://www.gartner.com/newsroom/id/2636073}.

\bibitem{arsGlobalInternet}
M.~Geuss.
\newblock {Satellite Internet: meet the hip new investment for Richard Branson,
  Elon Musk}.
\newblock \url{http://goo.gl/D0THPj}.

\bibitem{ha2013impact}
K.~Ha, P.~Pillai, G.~Lewis, S.~Simanta, S.~Clinch, N.~Davies, and
  M.~Satyanarayanan.
\newblock {The Impact of Mobile Multimedia Applications on Data Center
  Consolidation}.
\newblock {\em IC2E}, 2013.

\bibitem{habib}
M.~A. Habib and M.~Abrams.
\newblock {Analysis of Sources of Latency in Downloading Web Pages}.
\newblock {\em WEBNET}, 2000.

\bibitem{ilya-latency}
{Ilya Grigorik (Google)}.
\newblock {Latency: The New Web Performance Bottleneck}.
\newblock \url{http://goo.gl/djXp3}.

\bibitem{fbDrone}
I.~Lapowsky.
\newblock {Facebook Lays Out Its Roadmap for Creating Internet-Connected
  Drones}.
\newblock \url{http://goo.gl/HupzK2}.

\bibitem{chi-ny}
G.~Laughlin, A.~Aguirre, and J.~Grundfest.
\newblock {Information Transmission between Financial Markets in Chicago and
  New York}.
\newblock {\em Financial Review}, 2014.

\bibitem{amazonLatency}
J.~Liddle.
\newblock {Amazon Found Every 100ms of Latency Cost Them 1\% in Sales}.
\newblock \url{http://goo.gl/BUJgV}.

\bibitem{jumpstart}
D.~Liu, M.~Allman, S.~Jin, and L.~Wang.
\newblock Congestion control without a startup phase.
\newblock In {\em PFLDnet}, 2007.

\bibitem{rc3}
R.~Mittal, J.~Sherry, S.~Ratnasamy, and S.~Shenker.
\newblock Recursively cautious congestion control.
\newblock In {\em NSDI}. USENIX, 2014.

\bibitem{wolfgang}
W.~M{\"u}hlbauer, S.~Uhlig, A.~Feldmann, O.~Maennel, B.~Quoitin, and B.~Fu.
\newblock {Impact of Routing Parameters on Route Diversity and Path Inflation}.
\newblock {\em Computer Networks}, 2010.

\bibitem{apip}
D.~Naylor, M.~K. Mukerjee, and P.~Steenkiste.
\newblock Balancing accountability and privacy in the network.
\newblock In {\em SIGCOMM}. ACM, 2014.

\bibitem{seasCable}
NEC.
\newblock {SEA-US: Global Consortium to Build Cable System Connecting
  Indonesia, the Philippines, and the United States}.
\newblock \url{http://goo.gl/ZOV3qa}.

\bibitem{arcticCable}
A.~Nordrum.
\newblock {Fiber optics for the far North [News]}.
\newblock {\em Spectrum, IEEE}, 2015.

\bibitem{gamingLatency}
L.~Pantel and L.~C. Wolf.
\newblock {On the Impact of Delay on Real-Time Multiplayer Games}.
\newblock {\em NOSSDAV}, 2002.

\bibitem{patterson2004latency}
D.~A. Patterson.
\newblock {Latency Lags Bandwidth}.
\newblock {\em Communications of the ACM}, 2004.

\bibitem{onetunnel}
S.~Peter, U.~Javed, Q.~Zhang, D.~Woos, T.~Anderson, and A.~Krishnamurthy.
\newblock One tunnel is (often) enough.
\newblock In {\em SIGCOMM}. ACM, 2014.

\bibitem{transAtlantic}
M.~Price.
\newblock {Enduring Quest for Superfast Trading Takes to the Air}.
\newblock \url{http://goo.gl/JKpMX}.

\bibitem{fastopen}
S.~Radhakrishnan, Y.~Cheng, J.~Chu, A.~Jain, and B.~Raghavan.
\newblock {TCP Fast Open}.
\newblock {\em {CoNEXT}}, 2011.

\bibitem{routestable}
J.~Rexford, J.~Wang, Z.~Xiao, and Y.~Zhang.
\newblock {BGP Routing Stability of Popular Destinations}.
\newblock {\em ACM SIGCOMM Workshop on Internet Measurment}, 2002.

\bibitem{singla14hotnets}
A.~Singla, B.~Chandrasekaran, P.~B. Godfrey, and B.~Maggs.
\newblock {The Internet at the Speed of Light}.
\newblock In {\em HotNets}. ACM, 2014.

\bibitem{sundaresan}
S.~Sundaresan, N.~Magharei, N.~Feamster, and R.~Teixeira.
\newblock {Measuring and Mitigating Web Performance Bottlenecks in Broadband
  Access Networks}.
\newblock {\em IMC}, 2013.

\bibitem{tahtLatency}
D.~T\"{a}ht.
\newblock {On Reducing Latencies Below the Perceptible}.
\newblock {\em {Workshop on Reducing Internet Latency}}, 2013.

\bibitem{vulimiri}
A.~Vulimiri, P.~B. Godfrey, R.~Mittal, J.~Sherry, S.~Ratnasamy, and S.~Shenker.
\newblock {Low Latency via Redundancy}.
\newblock {\em {CoNEXT}}, 2013.

\bibitem{wprof}
X.~S. Wang, A.~Balasubramanian, A.~Krishnamurthy, and D.~Wetherall.
\newblock {Demystify Page Load Performance with WProf}.
\newblock {\em NSDI}, 2013.

\bibitem{wang}
Z.~Wang.
\newblock {Speeding Up Mobile Browsers without Infrastructure Support}.
\newblock Master's thesis, Duke University, 2012.

\bibitem{vrImmersion}
F.~Zheng, T.~Whitted, A.~Lastra, P.~Lincoln, A.~State, A.~Maimone, and
  H.~Fuchs.
\newblock Minimizing latency for augmented reality displays: Frames considered
  harmful.
\newblock In {\em ISMAR}. IEEE, 2014.

\bibitem{asap}
W.~Zhou, Q.~Li, M.~Caesar, and P.~B. Godfrey.
\newblock {ASAP}: A low-latency transport layer.
\newblock {\em {CoNEXT}}, 2011.

\end{thebibliography}
%}
}

%\vspace{4pt}
%\input{datapolicy}
%\vspace{4pt}
%\input{appendix}

\end{document}